\documentclass[twocolumn]{aastex62}

\newcommand\harm{\textit{HARM${^2}$}}
\def\vir{\texttt{Vir}}
\def\hii{H~\textsc{ii}}
\def\harm{\textit{HARM${^2}$}}
\def\orion{\textsc{Orion2}}

\def\hrtofb{\texttt{TurbRad+OFB}}
\def\hrtof{\texttt{TurbRad+OF}}
\def\hrt{\texttt{TurbRad}}

\usepackage[normalem]{ulem}

\newcommand{\black}[1]{{\textcolor{black}{#1}}}

\newcommand{\add}[1]{{\black{#1}}}

\usepackage{subfigure}

\received{April 17, 2020}
\revised{May 29, 2020}
\accepted{June 7, 2020}
\submitjournal{ApJ}

\shorttitle{Massive Star Formation with Outflows, Radiation Pressure, and Magnetic Fields}
\shortauthors{Rosen \& Krumholz}

\begin{document}

\title{The Role of Outflows, Radiation Pressure, and Magnetic Fields in Massive Star Formation}

\correspondingauthor{Anna Rosen}
\email{anna.rosen@cfa.harvard.edu}

\author[0000-0003-4423-0660]{Anna L. Rosen}
\altaffiliation{NASA Einstein Fellow}
\altaffiliation{ITC Fellow}
\affiliation{Center for Astrophysics $|$ Harvard \& Smithsonian, 60 Garden St, Cambridge, MA 02138, USA}

\author[0000-0003-3893-854X]{Mark R. Krumholz}
\affiliation{Research School of Astronomy and Astrophysics, Australian National University, Canberra, ACT 2611 Australia}
\affiliation{ARC Centre of Excellence for Astronomy in Three Dimensions (ASTRO-3D), Canberra, ACT 2611 Australia}

%\date{Accepted 2020 June 7}
%\maketitle

\begin{abstract}
Stellar feedback in the form of radiation pressure and magnetically-driven collimated outflows may limit the maximum mass that a star can achieve and affect the star-formation efficiency of massive pre-stellar cores. Here we present a series of 3D adaptive mesh refinement radiation-magnetohydrodynamic simulations of the collapse of initially turbulent, massive pre-stellar cores. Our simulations include radiative feedback from both the direct stellar and dust-reprocessed radiation fields, and collimated outflow feedback from the accreting stars. We find that protostellar outflows punches holes in the dusty circumstellar gas along the star's polar directions, thereby increasing the size of optically thin regions through which radiation can escape. Precession of the outflows as the star's spin axis changes due to the  turbulent accretion flow further broadens the outflow, and causes more material to be entrained. Additionally, the presence of magnetic fields in the entrained material leads to broader entrained outflows that escape the core. We compare the injected and entrained outflow properties and find that the entrained outflow mass is a factor of $\sim$3 larger than the injected mass and the momentum and energy contained in the entrained material are $\sim$25\% and $\sim$5\% of the injected momentum and energy, respectively. As a result, we find that, when one includes both outflows and radiation pressure, the former are a much more effective and important feedback mechanism, even for massive stars with significant radiative outputs.
\end{abstract}


\section{Introduction}

Massive stars ($M_{\rm \star} \gtrsim 8 \, M_{\rm \odot}$) are rare, representing only $\sim$1\% of the stellar population by number, yet they dominate energy injection into the interstellar medium (ISM) in star-forming galaxies because of their strong radiation fields, fast stellar winds, and supernova explosions. During their formation, massive protostars launch collimated bi-polar outflows that are qualitatively similar, but much more powerful, than those produced by low-mass protostars \citep{Bontemps1996a, Maud2015b, Bally2016a}. These similarities suggest that the driving mechanisms for such outflows originate from the same source, thereby providing evidence that low- and high-mass stars form in a similar fashion \citep{Maud2015b}. 

Massive stars form in dense ($\sim10^4-10^6 \; \rm cm^{-3}$), cold ($\sim 10$ K), turbulent, and magnetized gas within giant molecular clouds and giant massive filaments \citep[see detailed review by ][]{Rosen2020a}. One of the key signatures of massive star formation, when the protostars are heavily embedded, are molecular outflows \citep[e.g., ][]{Maud2015b, Pillai2019a}. These entrained outflows likely originate from collimated jets that are magnetically launched via the star-disk interaction \citep{Shu1988a, Pelletier1992a, Bontemps1996a, Maud2015b, Kolligan2018a}. As these jets leave the star-disk system they encounter molecular material, which they sweep up, and this material can potentially be ejected from star-forming environments leading to low star formation efficiencies \citep[SFEs,][]{Cunningham2011a, Maud2015a, Kuiper2016a, Staff2019a}. Additionally, the momentum injected by outflows may limit accretion onto the massive star, potentially setting an upper mass limit to the stellar initial mass function (IMF). Therefore, protostellar outflows are an important form of stellar feedback -- the injection of momentum and energy into the surrounding ISM by young stars -- during the star formation process. 

Entrained molecular outflows from massive protostars are often observed via molecular line tracers (e.g., SiO, CS, or CO), methanol or water masers, and non-thermal synchrotron emission. They typically show complex orientations and morphologies with multiple shocks and broad density structures \citep[e.g.,][]{deVilliers2014a, Maud2015b, Brogan2018a, Avison2019a, Gieser2019a, Sanna2019a}. Observations of outflows from both low- and high-mass protostars indicate that the outflow and infall motions occur simultaneously and are closely linked. Typically, the outflow mass-loss rates are inferred to be 10-30\% of the accretion rate onto the star. The outflow mass rate can also be highly variable, likely due to episodic accretion \citep{Bachiller1996a, Matzner2000a, Burns2020a, Nony2020a}.

Significant theoretical attention has focused primarily on the role that radiation pressure plays in massive star formation because massive stars have short Kelvin-Helmholtz time-scales (the time required for a star to radiate away its gravitational binding energy) and  attain their main-sequence luminosities while they are still actively accreting \citep{Palla1991a, Palla1992a, Behrend2001a, Hosokawa2009a, Krumholz2009a, Kuiper2011a, Rosen2016a}. The radiation pressure associated with their high luminosities can oppose gravity and halt accretion. However, the rate of momentum deposition attributed to the direct radiation pressure from stars (i.e., first absorption of the stellar radiation by interstellar dust), $\dot{p}_{\rm rad} = \frac{L_{\rm \star}}{c}$ where $L_{\rm \star}$ is the massive protostar's stellar luminosity, given by
\begin{equation}
\dot{p}_{\rm rad} = 2 \times 10^{-3}  \left(\frac{L_{\rm \star}}{10^5 \, L_{\odot}}\right) \rm{M_{\odot} \; km \; s^{-1} \; yr^{-1}},
\end{equation}
\noindent
is typically lower than the rate of momentum deposition from protostellar outflows, $\dot{p}_{\rm OF} = \dot{M}_{\rm OF} v_{\rm OF}$ where $\dot{M}_{\rm OF}$ is the outflow mass-loss rate and $v_{\rm OF}$ is the outflow velocity. This quantity is given by 

\begin{eqnarray}
\dot{p}_{\rm OF} &=&  4.5 \times 10^{-3}  \left( \frac{f_w}{0.2}\right) \left( \frac{f_k}{0.3}\right)  \left(\frac{M_{\rm \star}}{30\, M_{\odot}}\right)^{\frac{1}{2}} 
	\left(\frac{R_{\rm \star}}{10\, R_{\odot}}\right)^{\frac{1}{2}} \nonumber \\
 & & \times \left(\frac{\dot{M}_{\rm \star}}{10^{-4}\, \rm{M_{\odot} \; yr^{-1}}} \right) \rm{M_{\odot} \; km \; s^{-1} \; yr^{-1}}
 \label{eq:mdot_of_scaling}
\end{eqnarray}
\noindent
where we parameterize the outflow mass flux and velocity as $\dot{M}_{\rm OF} = f_w \dot{M}_{\rm acc}$ and $v_{\rm OF} = f_k v_{\rm kep}$, where $\dot{M}_{\rm acc}$ is the accretion rate, $f_w$ is the fraction of the accreted mass that is lost to outflows, $v_{\rm kep} = \sqrt{G M_{\rm \star}/R_{\rm \star}}$ is the Keplerian velocity of a star with mass $M_{\rm \star}$ and radius $R_{\rm \star}$, and $f_{\rm k}$ is the fraction of the Keplerian velocity  with which the outflow is launched \citep{Matzner2000a}. The values to which we have scaled in \autoref{eq:mdot_of_scaling} are typical for massive protostars ($M_\star$, $R_\star$, and $\dot{M}_\star$) and magnetocentrifugal wind models ($f_w$ and $f_k$). Therefore, momentum feedback from protostellar outflows is a non-negligible feedback mechanism in massive star formation that must be included when studying how feedback limits accretion onto massive stars.

Previous numerical studies have studied the effect outflows play in massive star formation in addition to radiation pressure. \citet{Cunningham2011a} performed a series of 3D adaptive mesh refinement (AMR) radiation-hydrodynamic (RHD) simulations with radiative and outflow feedback, where they modeled the dust reprocessed radiation field and used an \textit{ad-hoc} prescription to treat the stellar radiation field from stars. In agreement with \citet{Krumholz2005c}, they found that outflows evacuate polar cavities of reduced optical depth through the turbulent, ambient core. This effect enhances the radiative flux in the poleward direction so that radiative heating and the outward radiative force is diminished. Likewise, \citet{Kuiper2015a, Kuiper2016a} performed cylindrically symmetric RHD simulations in which the star is held fixed (i.e., outflow launching location and outflow direction remained constant) but they include a hybrid treatment of radiation pressure that properly accounts for both the  stellar  and dust reprocessed radiation fields inherent to massive star formation. They found that outflows open a bipolar cavity extending to the  outer edge of the protostellar core from which the simulation begins. The opening angles of the outflows and the amount of mass entrained and ejected from the core both increase with time. Additionally, they find that the importance of feedback from outflows depends on the amount of mass injected into the outflows at the point where they are launched, suggesting that outflows with a larger mass flux yields a star formation efficiency (SFE) from 50\% in the case of very weak outflows to as low as 20\% for very strong outflows. 

These studies concluded that protostellar outflows makes radiation pressure less significant in massive star formation and causes molecular material to be ejected from the core leading to low SFEs. However, the simulations by \citet{Kuiper2015a, Kuiper2016a} model the collapse of a laminar core with the star held fixed, so the outflow launching direction remains fixed. In reality, the accretion flow onto a massive protostar is chaotic because the accreted core material is turbulent, which in turn will cause the star's spin axis to precess. This effect should make the outflow launching direction highly variable, possibly leading to multiple outflows as are commonly observed in massive star forming regions \citep[e.g.,][]{Gieser2019a, Avison2019a}.  Additionally, magnetic fields are also not included in the \citealt{Kuiper2015a, Kuiper2016a} and \citealt{Cunningham2011a} simulations, and they will likely affect the outflow structure and ejection of material. 

In this paper, we investigate these effects by performing 3D radiation-magnetohydrodynamics (RMHD) numerical simulations of the collapse of magnetized and unmagnetized turbulent massive prestellar cores into massive stellar systems, including both radiative and outflow feedback. This paper is organized as follows: we describe our numerical methodology and simulation design in Section~\ref{sec:meth}; we present and discuss our results in Sections~\ref{sec:results} and \ref{sec:disc}, respectively; finally, we conclude in Section~\ref{sec:conc}.

\section{Numerical Method}
\label{sec:meth}

In this paper, we simulate the gravitational collapse of isolated magnetized and non-magnetized turbulent massive pre-stellar cores with the \orion\  adaptive mesh refinement (AMR) code. \orion\ includes MHD \citep{Li2012a}, radiative transfer \citep{Krumholz2007a, Shestakov2008a, Rosen2017a}, self-gravity \citep{truelove1998a}, and Lagrangian accreting sink particles \citep{Krumholz2004a} that include a protostellar evolution model used to represent them as radiating protostars \citep{Offner2009a} coupled to a sub-grid prescription to model stellar feedback from protostellar outflows \citep{Cunningham2011a}. We describe the equations solved by our code in Section \ref{sec:numerics}, our stellar radiation and outflow feedback prescriptions in Section \ref{sec:feedback}, and the initial and boundary conditions, including our refinement and sink creation requirements, for our simulations in Section \ref{sec:ics}. %and our refinement criteria and sink creation requirements in Section \ref{sec:sink}.

\subsection{Evolution Equations}
\label{sec:numerics}
The full gravito-RMHD equations solved by \orion\ for the simulations that describe the dynamics of the fluid-sink (star) particle system presented in this work are

\begin{eqnarray}
\label{eqn:com}
\frac{\partial \rho}{\partial t} & = & -\mathbf{\nabla} \cdot \left( \rho \mathbf{v} \right) - \sum_{i}  \dot{M}_{a,i} W_{a}(\mathbf{x} - \mathbf{x}_i)  \nonumber\\
	                                   &&+ \sum_i \dot{M}_{o,i} W_{o,i}(\mathbf{x} - \mathbf{x}_i) \\
%%%
\label{eqn:cop}
\frac{\partial \left(\rho \mathbf{v} \right)}{\partial t}  &=& -\mathbf{\nabla} \cdot \left( \rho \mathbf{v}  \bf{v} \right) - \mathbf{\nabla} \left( P + \frac{B^2}{8\pi}\right) + \frac{1}{4\pi} \mathbf{B} \cdot \mathbf{\nabla} \mathbf{B} \nonumber\\ 
	&& {} - \rho \mathbf{\nabla} \phi - \lambda \mathbf{\nabla} E_{\rm R}  - \sum_i \dot{\mathbf{p}}_{a,i}W_{a}(\mathbf{x} - \mathbf{x}_i) \nonumber\\
	&& + \sum_i \dot{\mathbf{p}}_{\rm rad, \it i} + \sum_i \dot{\mathbf{p}}_{o, \it i} W_{o,i}(\mathbf{x} - \mathbf{x}_i) 
\\
%%%
\label{eqn:coe}
\frac{\partial \left( \rho e\right)}{\partial t} &=& - \mathbf{\nabla} \cdot \left[ (\rho e + P + \frac{B^2}{8 \pi})\textbf{v}   
	- \frac{1}{4\pi} \mathbf{B} (\bf{v} \cdot \mathbf{B}) \right]\nonumber \\
&& - \rho \mathbf{v} \cdot \mathbf{\nabla} \phi - \kappa_{\rm 0P} \rho(4\pi B_{P} - cE_{\rm R} ) \nonumber \\ 
	&&+ \lambda \left( 2 \frac{\kappa_{\rm 0P}}{\kappa_{\rm 0R}} - 1 \right) \mathbf{v} \cdot \mathbf{\nabla} E_{\rm R} \nonumber \\
	&& - \left(\frac{\rho}{m_{\rm p}} \right)^2 \Lambda(T_{\rm g}) -  \sum_i \dot{\varepsilon}_{a,i} W_{a,i}(\mathbf{x} - \mathbf{x}_i) \nonumber \\
	&&+ \sum_i \dot{\mathbf{\varepsilon}}_{\rm rad, \it i} + \sum_i \dot{\mathbf{\varepsilon}}_{o,i} W_{a}(\mathbf{x} - \mathbf{x}_i) 
\\
%%%
\label{eqn:coer}
\frac{\partial E_{\rm R}}{\partial t} &=& \mathbf{\nabla} \cdot \left( \frac{c \lambda}{\kappa_{\rm 0R} \rho} \mathbf{\nabla} E_{\rm R} \right) + \kappa_{\rm 0P} \rho \left(4\pi B_P - c E_{\rm R} \right) \nonumber \\
	&&- \lambda \left( 2 \frac{\kappa_{0 \rm P}}{\kappa_{0 R}} - 1\right) \mathbf{v} \cdot \nabla E_{\rm R} 
	- \nabla \cdot \left( \frac{3 - R_2}{2} \mathbf{v} E_{\rm R}\right) \nonumber \\ 
	& & +  \left(\frac{\rho}{m_{\rm p}} \right)^2 \Lambda(T_{\rm g}) 
%%% 
\\
\label{eqn:induct}
\frac{\partial \bf{B}}{\partial t}&& = \mathbf{\nabla} \times (\bf{v} \times \bf{B})
\\
\label{eqn:mdot}
\frac{d M_{\rm i}}{dt} &&= \dot{M}_{a,i} \\
\label{eqn:dvi}
\frac{d \bf{x}_{i}}{dt} &&=\frac{\bf{p}_{i}}{M_{\rm i}} \\
\label{eqn:dpi}
\frac{d\bf{p}_i}{dt} &&= -M_i \nabla \phi + \dot{\bf{p}}_{a,i} \\
\label{eqn:pois}
\nabla^2 \phi &&= 4 \pi G \left[ \rho + \sum_i M_i \delta(\bf{x} - \bf{x}_i)\right].
\end{eqnarray}

\noindent
Here $\rho$ is the density, $\rho \mathbf{v}$  is the momentum density, $\rho e$ is the total internal plus kinetic gas energy density, $E_{\rm R}$ is the radiation energy density in the rest frame of the computational domain, $\mathbf{B}$ is the magnetic field, and $\phi$ is the gravitational potential. Equations \ref{eqn:com}-\ref{eqn:coer} describe conservation of gas mass, gas momentum, gas total energy, and radiation total energy, respectively. They include terms describing the exchange of these quantities with the star particles, and exchange of energy and momenta between radiation, magnetic fields, and gas. Equation~\ref{eqn:induct} is the induction equation that describes the time evolution of the magnetic field in the ideal limit that assumes the magnetic field and fluid are well-coupled. \orion\ includes MHD using a constrained transport scheme \citep{Li2012a} that maintains $\mathbf{\nabla} \cdot \mathbf{B} = 0$ to machine accuracy.

We assume an ideal equation of state so that the gas pressure is 
\begin{equation}
P=\frac{\rho k_{\rm B}T}{\mu m_{\rm H}} = \left( \gamma-1\right) \rho e_{\rm T},
\end{equation}
where $T$ is the gas temperature, $\mu$ is the mean molecular weight,  $\gamma$ is the ratio of specific heats, and $e_{\rm T}$ is the thermal energy of the gas per unit mass. We take $\mu=2.33$  and $\gamma=5/3$ that is appropriate for molecular gas of solar composition at temperatures too low to excite the rotational levels of H$_2$. We assume the fluid is a mixture of gas and dust with a dust-to-gas ratio of 0.01. At the high densities that we are concerned with the dust will be thermally coupled to the gas, allowing us to assume that the dust and gas temperatures are the same. 

The radiation-specific quantities in Equations~\ref{eqn:coe}-\ref{eqn:coer} are the Planck- and Rosseland-mean opacities $\kappa_{\rm 0P}$ and $\kappa_{\rm 0R}$ computed in the frame co-moving with the gas,  the  blackbody function $B_{\rm P} = c a_{\rm R} T^4/(4\pi)$, a dimensionless number $\lambda$ called the flux limiter, and the Eddington factor $R_2$. The last two quantities originate from the (gray) flux limited diffusion (FLD) approximation (see \citet{Krumholz2007a}  and \citet{Rosen2016a} for more detail). Lastly, $\Lambda(T)$ describes the cooling rate by atomic lines and the continuum, which only becomes significant when $T \gtrsim 10^3$ K when dust  begins to sublime \citep{Cunningham2011a}.

We evolve the radiating (proto)star particles via Equations~\ref{eqn:mdot}-\ref{eqn:dpi}, indexed by the subscript $i$ in the above equations. These particles accrete nearby gas and interact with the fluid via gravity, radiation, and protostellar outflows. We describe the modeling of their feedback (i.e., the momentum and energy injected into the fluid)  associated with their radiation fields and outflows in Section~\ref{sec:feedback}, but note that the radiation and outflow specific terms in Equations~\ref{eqn:com}-\ref{eqn:coe} affiliated with star particles are denoted with the $\mathrm{rad}$ and $o$ subscripts, respectively. The star particles are characterized by their position $\mathbf{x}_{i}$, momentum $\mathbf{p}_i$, mass $M_i$, \add{angular momentum that describes the particle's spin axis $\mathbf{J}_i$,} and luminosity $\dot{\varepsilon}_{\mathrm{rad},i}$ \citep[e.g., integrated stellar spectrum from ] []{Lejeune1997a}, as determined by the protostellar evolution model described in \citet{Offner2009a}. They accrete mass, momentum, and energy from the computational grid at rates $\dot{M}_{a,i}$, $\dot{\mathbf{p}}_{a,i}$, and $\dot{\varepsilon}_{a,i}$. The distribution of these quantities over cells in the computational grid is described by a weighting kernel $W_{a}(\mathbf{x}-\mathbf{x}_i)$, which is non-zero only within 4 computational zones of each particle, following the algorithm of \citet{Krumholz2004a}. \add{We update the star particles' angular momentum and spin axis directions via the prescription described in \citet{Fielding2015a}.} The gravitational potential of the gas is advanced by solving Poisson's equation given by Equation~\ref{eqn:pois}, which includes contributions from both the fluid and star particles. 

For each simulation we begin with a base grid with volume (0.4 pc)$^3$ discretized by $128^3$ cells and allow for five levels of refinement, resulting in a maximum resolution of 20 au. As the simulation evolves, the AMR algorithm automatically adds and removes finer grids based on certain refinement criteria set by the user. We refine cells if they meet at least one of the following criteria: (1) any cell on the base level (i.e., level 0) that has a density equal to or greater than the core's edge density, so that the entire prestellar core is refined to level 1; (2) any cell where the density in the cell exceeds the Jeans density given by

\begin{equation}
\label{eqn:rhoj}
\rho_{\rm max,J} = \frac{\pi J^2_{\rm max} c_{\rm s}^2}{G \Delta x^2_l} \left( 1 + \frac{0.74}{\beta^2}\right),
\end{equation}
\noindent
where $c_{s}=\sqrt{kT/\mu m_{\rm{p}}}$ is the isothermal sound speed, $\Delta x_l$ is the cell size on level $l$, $\beta=8\pi \rho c_s^2/B^2$, and $J_{\rm max}$ is the maximum allowed number of Jeans lengths per cell, which we set to 1/8, following the MHD Truelove Criterion \citep{Myers2013a};  (3) any cell that is located within at least eight cells of a sink particle; and (4) any cell within which the radiation energy density gradient exceeds $\nabla E_{\rm R} > 0.15 E_{\rm R}/\Delta x_{l}$. %This procedure is repeated recursively on all levels after every two level updates. 

Star particles form when the Jeans condition for a Jeans number of $N_J$ = 0.25 is exceeded on the maximum AMR level following the resolution tests of \citet{Truelove1997a}. They are allowed to merge when two star particles pass within one accretion radius of each other if the smaller particle has a mass less than $0.04 \; M_{\rm \odot}$, the threshold that corresponds to the largest plausible mass at which second collapse occurs for the protostar \citep{Masunaga1998a, Masunaga2000a}. At masses lower than this value the protostar represents a hydrostatic core that is several au in size and will likely be accreted by the more massive star whereas larger mass protostars will have collapsed down to sizes of roughly several $R_{\rm \odot}$ and will unlikely merge with the nearby protostar.

\begin{table*}
	\begin{center}
	\caption{
	\label{tab:sim}
Simulation Parameters
}

	\begin{tabular}{ l c  c  c  c  c  c  c c}
	\hline
	\textbf{Run} & & \hrt\ & \hrtof\  & \hrtofb\ \\
	\\
	\hline
	\textbf{Physical Parameter}\\
	\hline
	\hline
	Core Mass [$\rm M_{\rm \odot}$] & $M_{\rm c}$ & 150 & 150 & 150\\
	Core Radius [pc] & $R_{\rm c}$ & 0.1 & 0.1 & 0.1\\
	Surface Density [$\rm g \, cm^{-2}$] & $\Sigma$ & 1 & 1 &1\\
	Temperature [K] & $T_{\rm c}$ & 20 & 20 & 20 \\
	Mean Density [$\rm 10^{-18}\, g \, cm^{-3}$] & $\bar{\rm \rho}_{\rm cl}$ & $2.4$ & $2.4$ & $2.4$\\
	Mean Free-fall Time  [kyr] & $t_{\rm ff}$ & 42.8 & 42.8 & 42.8\\
	Power Law Index & $k_{\rm \rho}$ & 1.5 & 1.5 & 1.5\\
	Velocity Dispersion [$\rm km \, s^{-1} $] \tablenotemark{a} & $\sigma_{\rm 1D}$ & 1.2 & 1.2 & 1.2\\
	Mass-to-flux ratio & $\mu_{\rm \phi}$ & $\infty$ & $\infty$ & 2 \\
	Magnetic Field Strength [$\rm  mG $] & $B_z$  & 0 & 0 & 0.81 \\
	\hline
	\\
	\textbf{Numerical Parameter}\\
	\hline
	\hline
	Rad. Feedback? & & Yes & Yes  & Yes\\
	Outflows? & & No & Yes & Yes \\
	EOS Index \tablenotemark{b} & $n$ & 5/3  & 5/3 & 5/3\\
	Domain Length [pc] & $L_{\rm box}$ & 0.4 & 0.4  & 0.4\\
	Base Grid Cells & $N_{\rm 0}$ & $128^3$ & $128^3$  & $128^3$\\ 
	Maximum Level & $l_{\rm max}$ & 5  & 5 & 5\\
	Minimum Cell Size [au] & $\Delta x_{\rm l_{\rm max}}$ &  20 & 20 & 20 \\
	Jeans Length Refinement &$J_{\rm max}$ & 0.125 & 0.125 & 0.125 \\
	$E_{\rm R}$ Gradient Refinement & $E_{\rm R}/\Delta x$ & 0.15 & 0.15  & 0.15\\ 
	Accretion radius [au] && 80 & 80 & 80 \\ 
	\hline
	\\
	\textbf{Simulation Outcome} \\
	\hline
	\hline
	Simulation Time [$t_{\rm ff}$] & & 0.95 & 0.95  & 1.36 \\
	Massive Star Mass [$\rm M_{\odot}$] & & 51.97 & 35.22  & 33.64\\
	Number of Sinks \tablenotemark{c} && 18 & 14 & 1\\
	\end{tabular}
	\tablenotetext{a}{ Volume-weighted.} 
	\tablenotetext{b}{ Equation of state: $P \propto \rho^{n}$.} 
	\tablenotetext{c}{Final number of sinks with masses greater than 0.04 $M_{\rm \odot}$.}
	\end{center}
\end{table*}

\subsection{Stellar Radiation and Collimated Outflow Feedback Modeling}
\label{sec:feedback}
Each star produces a (direct) stellar radiation field and collimated protostellar outflows that injects energy ($\varepsilon$) and momentum ($\bf{p}$) into the fluid at a rate per unit volume $\dot{\varepsilon}_{{\rm rad},i}$, $\dot{\varepsilon}_{o,i}$, $\dot{\mathbf{p}}_{{\rm rad},i}$, and $\dot{\mathbf{p}}_{o,i}$, where quantities subscripted by  $\mathrm{rad}$ and $o$ denote feedback from radiation and outflows, respectively. We use the multi-frequency Hybrid Adaptive Ray-Moment Method (\harm) described in \citet{Rosen2016a}  and \citet{Rosen2017a} to treat both the direct (stellar) and indirect (dust-reprocessed) radiation fields. This method combines direct solution of the frequency-dependent radiative transfer equation along long characteristics launched from stars to treat the direct stellar radiation field, including contributions from the accretion luminosity,
\begin{equation}
L_{\rm acc} = f_{\rm rad} \frac{G M_{ \star} \dot{M}_{\star}}{R_{\rm \star}}
\end{equation}
with a gray FLD method to treat the (indirect) radiation field produced by thermal emission from dust \citep{Krumholz2007a, Rosen2016a, Rosen2017a}. Here $f_{\rm rad}$ is the fraction of the gravitational potential energy of the accretion flow that is converted to radiation and we take $f_{\rm rad}=3/4$ \citep{Offner2009a}, and $M_{\rm \star}$ and $R_{\rm \star}$ are the star's mass and radius, respectively. We use the frequency-dependent stellar spectra and dust opacities from \citet{Lejeune1997a} and \citet{Weingartner2001a}, respectively, and divide the stellar spectrum and dust opacities into 10 frequency bins (e.g., see Figure~1 in \citet{Rosen2016a}). We refer the reader to \citet{Krumholz2007a} and \citet{Rosen2016a, Rosen2017a} for a detailed description of our treatment of the direct and indirect radiation pressures modeled in this work.

Proper modeling of the magnetic launching of collimated protostellar outflows requires sufficiently high-resolution \citep[e.g., sub-au;][]{Kolligan2018a}, which is prohibitively computationally expensive for the simulations presented in this work. Instead, we adopt a sub-grid prescription for launching outflows from stars based on the protostellar outflow model of  \citet{Matzner2000a}, first implemented by \citet{Cunningham2011a}. In this model, the outflows are described by a collimation angle, $\theta_c$, and launching fraction, $f_w$. This algorithm has the advantage that it can represent either a X-wind \citep{Shu1988a} or disk wind \citep{Pelletier1992a} model. For the simulations presented in this work we adopt $\theta_c=0.01$ and $f_w=0.21$, that assumes 21\% of the accreted material is ejected in the outflows, and inject the outflows in the 8 nearest zones to the star with the weighting kernel $W_{o,i}(\mathbf{x} - \mathbf{x}_i)$ described in \citet{Cunningham2011a} . The outflows are launched along the star's angular momentum (spin) axis at a fraction $f_k = 0.3$ of the Keplerian velocity, such that their velocity is $v_o =f_k \sqrt{G M_{\rm \star}/ R_{\rm \star}}$. These parameter values are chosen to match observations of outflow momentum observed in low- and high-mass star formation \citep{Cunningham2011a}.

The outflows inject mass, momentum $\dot{p}_o= \dot{M}_o v_o$, thermal energy $\dot{E}_{T,o} = \frac{\dot{M} kT_o}{\mu_w m_{\rm p}(\gamma-1)}$ where $T_{o}$ is the outflow gas temperature, and kinetic energy $\dot{E}_{k,o} = \frac{1}{2} \dot{M}_w v_o^2$ into the surrounding gas. To trace the outflow material we add a passively advected scalar to represent the outflow gas that is injected and we set $T_o$ equal to the protostar's surface temperature and $\mu_w=1.27$, which is the mean molecular weight for a neutral gas of solar composition, since observations have shown that outflows from intermediate and massive protostars are predominately neutral \citep{Reiter2016a, Cesaroni2018a, Fedriani2019a}. When the massive (proto)star reaches a surface temperature $\gtrsim 10^4$~K we then assume the outflow is ionized and set $T_{\rm w} = 10^4$K. 

\begin{figure*}
\centerline{\includegraphics[trim=0.2cm 1.5cm 0.2cm 1.0cm,clip,width=1.0\textwidth]{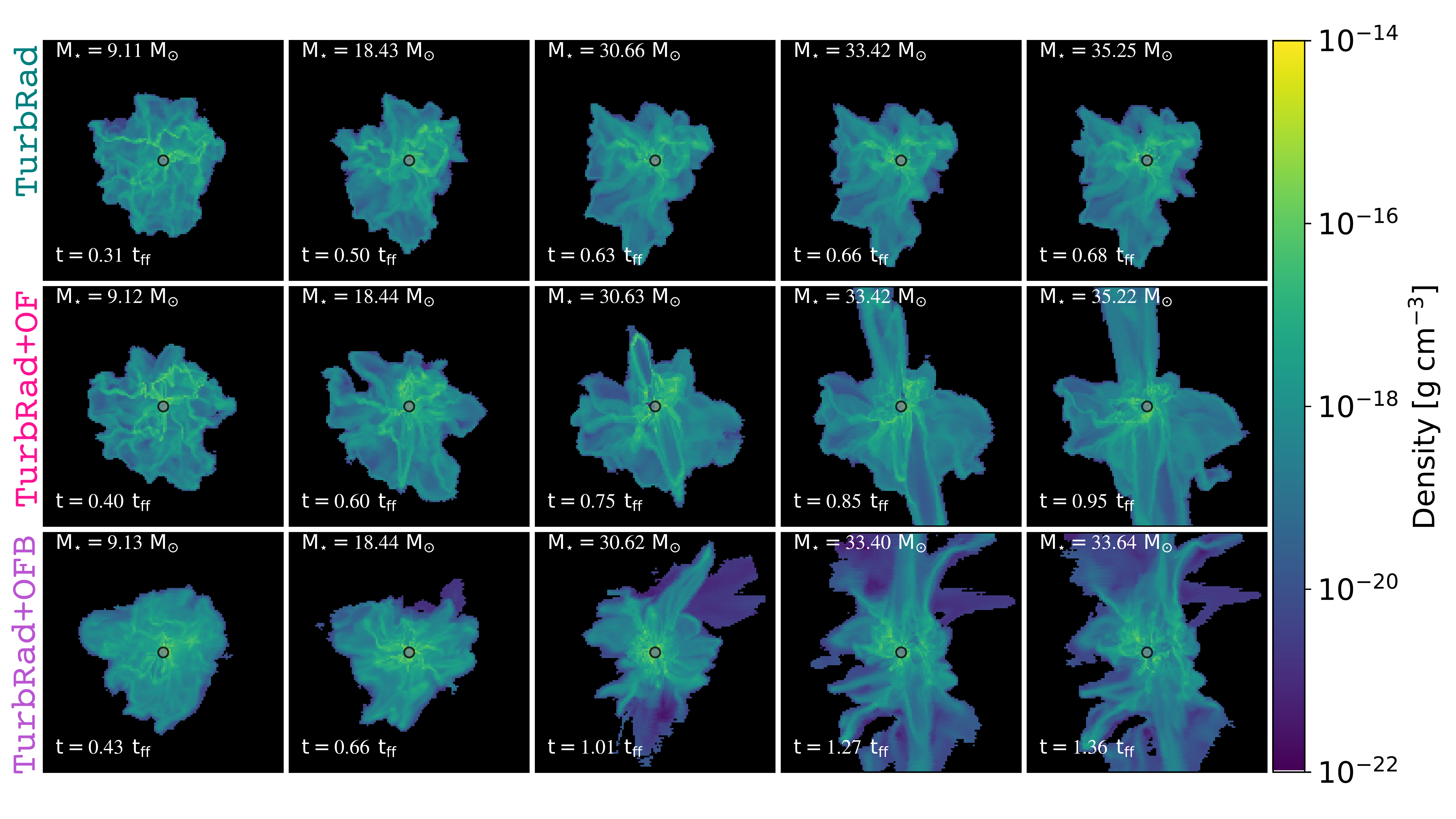}}%Turb_cloud_density_Both_multipanel_d500AUbig_Jstar.pdf}}
\caption{
\label{fig:density}
Density slices for runs \hrt\ (top row), \hrtof\ (middle row), and \hrtofb\ (bottom row). The most massive star is located at the center of each panel, as marked by the gray circle, and the slice is oriented such that its angular momentum axis points up, in order to highlight the density structure of the outflows. The primary stellar mass and the simulation time, in units of $t_{\rm ff}$, are shown in the bottom and top left corners of each panel, respectively. Each panel is (0.4 pc)$^2$.
}
\end{figure*}

\subsection{Initial and Boundary Conditions}
\label{sec:ics}
In this work, we perform three simulations of the collapse of turbulent, massive prestellar cores with feedback from stellar radiation and collimated outflows to determine how these feedback mechanisms effect the formation of massive stars: runs \hrt, \hrtof, and \hrtofb. Run \hrt\footnote{We note that run \hrt\ is the same run as \vir\ from \citet{Rosen2019a}} only includes radiative feedback, \hrtof\  includes radiative feedback and collimated protostellar outflows, and \hrtofb\ is identical to  \hrtof\ except that we also include magnetic fields. We use these simulations to compare how magnetic fields and feedback from collimated outflows affect massive star formation and the growth rate of massive stars.

Following \citet{Rosen2016a, Rosen2019a}, we begin with an isolated pre-stellar core of molecular gas and dust,  where we assume a dust-to-gas ratio of 0.01, with mass $M_{\rm c}=150 \; M_{\rm \odot}$, radius $R_{\rm c} = 0.1$ pc, and initial gas temperature of 20 K corresponding to a surface density of $\Sigma = M_{\rm c}/\pi R^2_{\rm c} = 1 \; \rm{g \; cm^{-2}}$ consistent with massive pre-stellar core densities and radii in extreme massive star forming environments \citep[\add{e.g.,}][]{Galvan2013a, Battersby2014a, Ginsburg2015a, Ginsburg2018a, Contreras2018a, Cao2019a}. The corresponding mean density of the core is $\bar{\rho} = 2.4 \times 10^{-18} \; \rm{g \; cm^{-3}}$ ($1.2 \times 10^6 \; \rm{H \; nuclei \; cm^{-3}}$) and its characteristic free-fall collapse time scale is $t_{\rm ff}\approx 42.6 \; \rm{kyr}$.  The core follows a density profile $\rho(r) \propto r^{- 3/2}$ in agreement with observations of massive cores at the $\sim$0.1 pc scale and clumps at the $\sim$1 pc scale that find values of $\kappa_{\rm \rho} = 1.5-2$ \citep[e.g.,][]{Caselli1995a, Beuther2002b, Mueller2002a, Beuther2007a, Zhang2009a, Longmore2011a, Butler2012a, Battersby2014a, Stutz2016a}. Each core is placed in the center of a 0.4 pc box that is filled with hot, diffuse gas with density $\rho_{\rm amb} = 0.01 \rho_{\rm edge}$ where $\rho_{\rm edge}$ is the density at the core boundary and temperature $T_{\rm amb} = 2000$ K so that the core is in thermal pressure balance with the ambient dust-free medium and we set the opacity of the ambient medium to zero. 

We explore the influence of magnetic fields on the collapse and outflow properties in run \hrtofb. In this run, we set the initial magnetic field to be uniform in the z direction: $\mathbf{B} = B_0 \hat{z}$. We choose $B_0= 0.81$ mG by selecting a mass-to-flux ratio $\mu_\Phi = M_{\rm c}/M_{\rm \phi} \simeq 2 \pi G^{1/2} M_{\rm c}/ \Phi = 2$, where $\Phi = \pi R^2_c B_0$ is the magnetic flux through the core, consistent with observed values of $\Phi \simeq$ 2-3 \citep{Crutcher2012a}. 

Observations of massive prestellar cores find that they contain supersonic turbulence and therefore we include supersonic turbulence by seeding the initial gas velocities ($v_x$, $v_y$, and $v_z$) with a velocity power spectrum, $P(k) \propto k^{-2}$ \citep{Padoan1999a, Boldyrev2002a, Cho2003a, Kowal2007a}. We include modes between $k_{\rm min} = 1$ to $k_{\rm max} = 256$ and take the turbulence mixture of gas to be 1/3 compressive and 2/3 solenoidal, consistent with the natural mixture of a 3D fluid \citep{Kowal2007a, Kowal2010a}. The onset of turbulence modifies the density and magnetic field distribution and we allow the turbulence to decay freely. For all runs, we use the same velocity perturbation power spectrum at initialization and a velocity dispersion of  $\sigma_{\rm 1D}= 1.2$ km/s corresponding to a $\alpha_{\rm vir} = 5 \sigma_{\rm 1D}^2 R_{\rm c}/G M_{\rm c}=1.1$ so that the core is roughly virialized. We allow the turbulence to decay, which is somewhat unrealistic. However, this simplification should have little effect on our results since the decay timescale, $\sim D/\sigma_{\rm 1D}$ where $D$ is the core diameter \citep{Goldreich1995a}, is  $\sim$0.16 Myr, which is much longer than the runtime for the simulations presented in this work.

\begin{figure*}
\centerline{\includegraphics[trim=0.2cm 1.25cm 0.2cm 1cm,clip,width=\textwidth]{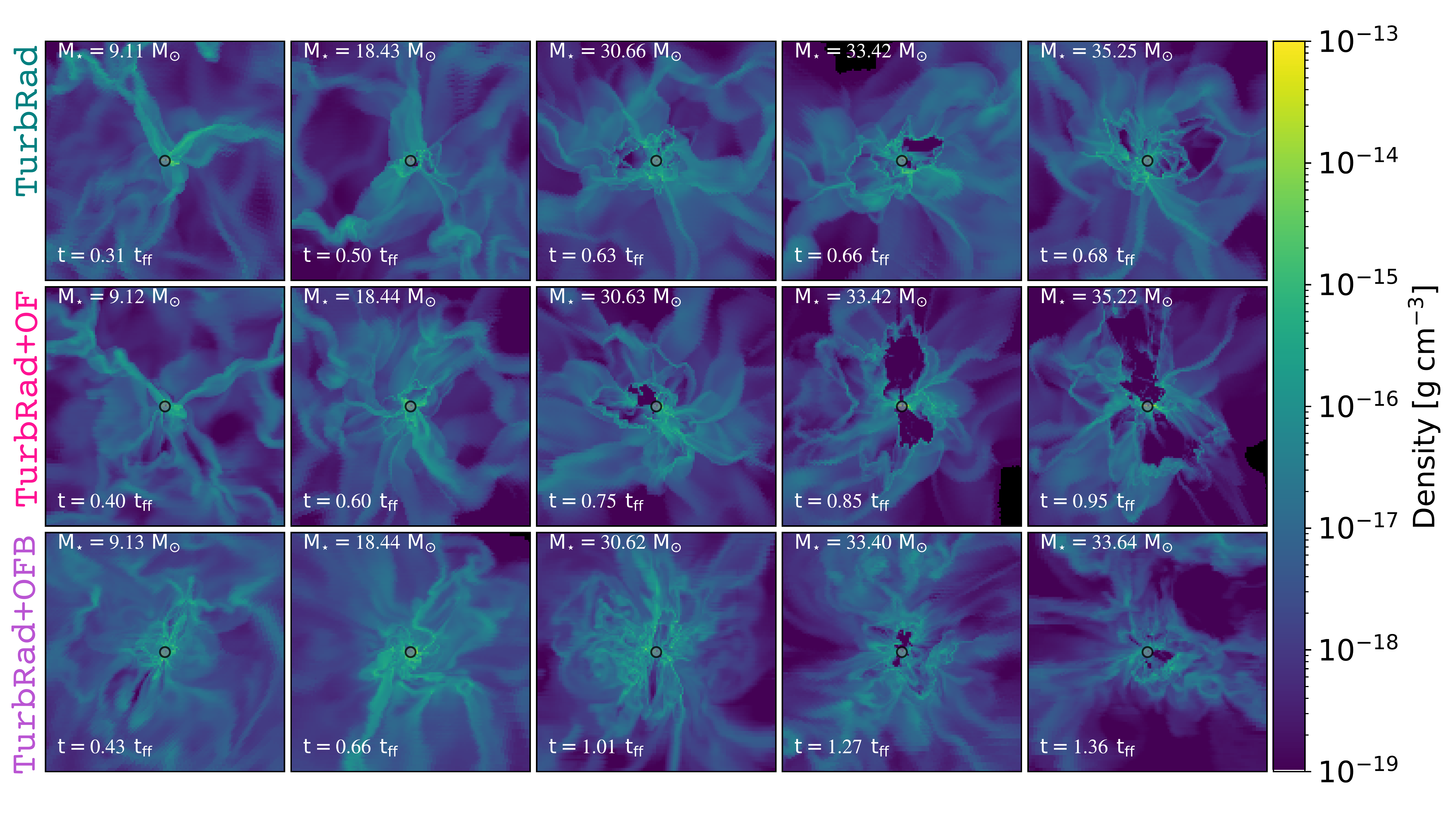}}%Turb_cloud_density_Both_multipanel_d500AUsmall_Jgas.pdf}}
\caption{
\label{fig:densityzoom}
Zoom-in density slices for runs \hrt\ (top row), \hrtof\ (middle row), and \hrtofb\ (bottom row). The most massive star is located at the center of each panel, as marked by the gray circle, and the slice is oriented such that the mass-weighted angular momentum axis of the gas within a radius of 250 au from the primary star points up in order to highlight the radiation pressure dominated bubbles that are perpendicular to the circumstellar gas due to the ``flashlight effect" as described in the text. The primary stellar mass and the simulation time, in units of $t_{\rm ff}$, are shown in the bottom and top left corners of each panel, respectively. Each panel is (0.1 pc)$^2$.
}
\end{figure*}

Our boundary conditions for the hydrodynamic, gravity, and radiation solvers are as follows. We impose outflow boundary conditions for the hydrodynamic update by setting the gradients of the hydrodynamic quantities $\left(\rho, \; \rho \bf{v}, \rho e \right)$ to be zero at the domain when advancing equations~\ref{eqn:com}-\ref{eqn:coe} \citep{Cunningham2011a, Myers2013a, Rosen2016a, Rosen2019a} and set the gravitational potential, $\phi$, to zero at all boundaries when solving Equation~\ref{eqn:pois} \citep{Myers2013a}. We do not expect this choice of boundary conditions for the gravitational potential to lead to any significant square artifacts near the domain boundaries since the core boundaries are far removed from the domain boundaries. Finally, for each radiation update, we impose Marshak boundary conditions that bathe the simulation volume with a blackbody radiation field equal to $E_0 = 1.21 \times 10^{-9} \rm{\; erg \; cm^{-3}}$ corresponding to a 20 K blackbody but we allow radiation generated within the simulation volume to escape freely \citep{Krumholz2009a, Cunningham2011a, Myers2013a, Rosen2016a, Rosen2019a}.

\section{Results}
\label{sec:results}
Here, we summarize the main results of our calculations. These simulations were run on the NASA supercomputer Pleiades located at NASA Ames. We use the \textsc{yt} package \citep{Turk2011a} to produce all the figures and quantitative analysis shown below. 

\begin{figure*}
\centerline{\includegraphics[trim=0.2cm 1.25cm 0.2cm 0.75cm,clip,width=1\textwidth]{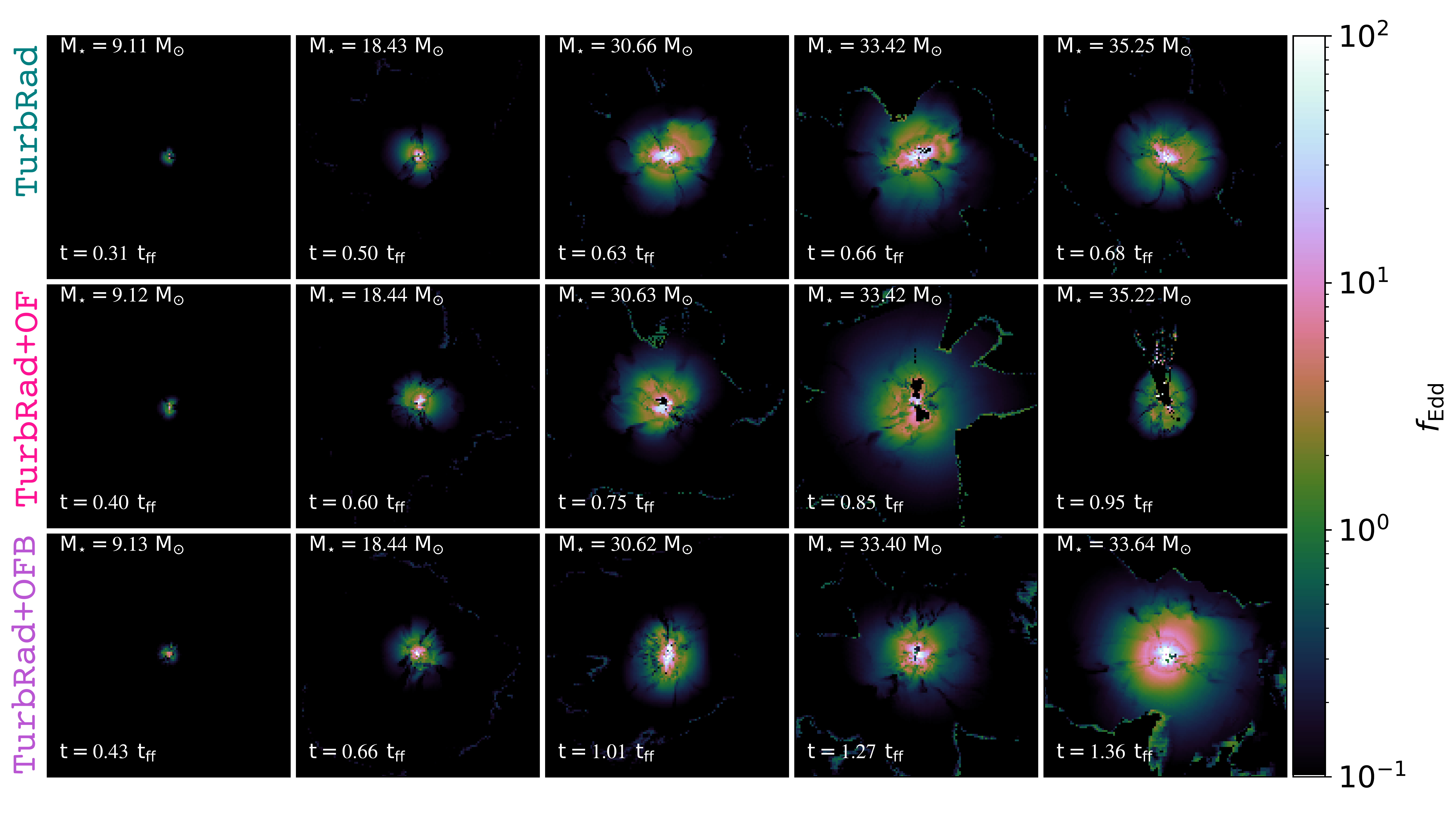}}%Turb_fedd_Both_multipanel_d500AUbig_Jgas.pdf}}
\caption{
\label{fig:fedd}
Slices of the Eddington ratio ($f_{\rm Edd} = f_{\rm rad}/f_{\rm grav}$) for runs \hrt\ (top row),  \hrtof\ (middle row), and \hrtofb\ (bottom row). The most massive star is located at the center of each panel and the slice is oriented such that the mass-weighted angular momentum axis of the gas within a radius of 250 au from the primary star points up. The primary stellar mass and the simulation time, in units of $t_{\rm ff}$, are shown in the bottom and top left corners of each panel, respectively. Each panel is (0.25 pc)$^2$.} 
\end{figure*}

\subsection{Density Structure}
Figure~\ref{fig:density} shows a series of density slices for runs \hrt\ (top row), \hrtof\ (middle row), and \hrtofb\ (bottom row) at the same primary (most massive) stellar mass. Each panel is oriented so that the primary star's angular momentum (spin) axis is pointing up and the center of each panel corresponds to the location of the primary star. Each slice covers an area of (0.4 pc)$^2$ with the primary star at the center. We choose this orientation for the density slices because the collimated outflows are injected along the direction of the stellar spin axis, and thus the overall outflow structure is predominantly along this axis. Comparison of the overall core density structure for runs \hrtofb\ and \hrtof\ show that the entrained outflows break out of the core when the star reaches a mass of $\sim 30 \; M_{\rm \odot}$ and that the outflows at breakout for run \hrtof\ are more collimated than those in run \hrtofb. The outflows in run \hrtofb\ are surrounded by a lower density envelope of material. We discuss the outflow structure and energetics in more detail in Section~\ref{sec:outflows}. Comparison of these simulations with run \hrt\ shows that the escaping outflows are a product of feedback from jets alone, because we do not see any outflow or bubble breakout from the core for run \hrt. 

Once the primary star in our simulations exceeds $\approx 30$ $M_\odot$, feedback from radiation pressure drives low-density radiation pressure dominated bubbles that expand away from the massive star. We show this in Figure~2, which shows zoom-in density slice plots for all three runs that cover an area of (0.1 pc)$^2$. This effect is commonly referred to as the ``flashlight effect'', in which optically thick circumstellar material pinches the radiation field and beams it into the polar directions, driving low-density cavities that expand away from the star \citep[e.g.,][]{Yorke2002a}. At equal primary stellar mass, the radiation pressure-dominated bubbles are more prominent in run \hrtof\ (middle row) as compared to run \hrt\ because outflows carve out low-density regions allowing the direct radiation pressure to be more effective at launching material at greater distances from the star. We find that these radiation pressure dominated low-density cavities are the least prominent in run \hrtofb. However, at equal times, the radiation pressure dominated bubbles are most prominent in run \hrt\ due to the faster mass growth and resulting larger luminosity of the primary star. Additionally, we find that  the low-density cavities that develop in run \hrtof\ are more collimated compared to the cavities in run \hrt\ at equal primary mass. %and equal times.

The expanding radiation pressure-dominated regions develop because material near the star becomes super-Eddington, as shown in Figure~\ref{fig:fedd}. As the central star gains mass, its luminosity increases, and the size of the super-Eddington regions near the star typically increase as well. %This occurs throughout the entirety of run \hrt\ and \hrtofb, but the super-Eddington region in run \hrtof\ eventually decreases at late times (e.g., see last panel of the middle row in Figure~\ref{fig:fedd}).
We note that  the super-Eddington region in run \hrtof\ eventually decreases at late times (e.g., see last panel of the middle row in Figure~\ref{fig:fedd}). This result for run \hrtof\ is likely due to the influence of outflows, which drive low-density channels near the star through which radiation can vent, an effect also seen by \citet{Cunningham2011a}, and predicted theoretically by \citet{Krumholz2005c}. %; we discuss this phenomenon in more detail in Section~\ref{sec:rad}. 

\begin{figure} 
\centerline{\includegraphics[trim=0.25cm 0.25cm 0.25cm 0.25cm,clip, width=1\columnwidth]{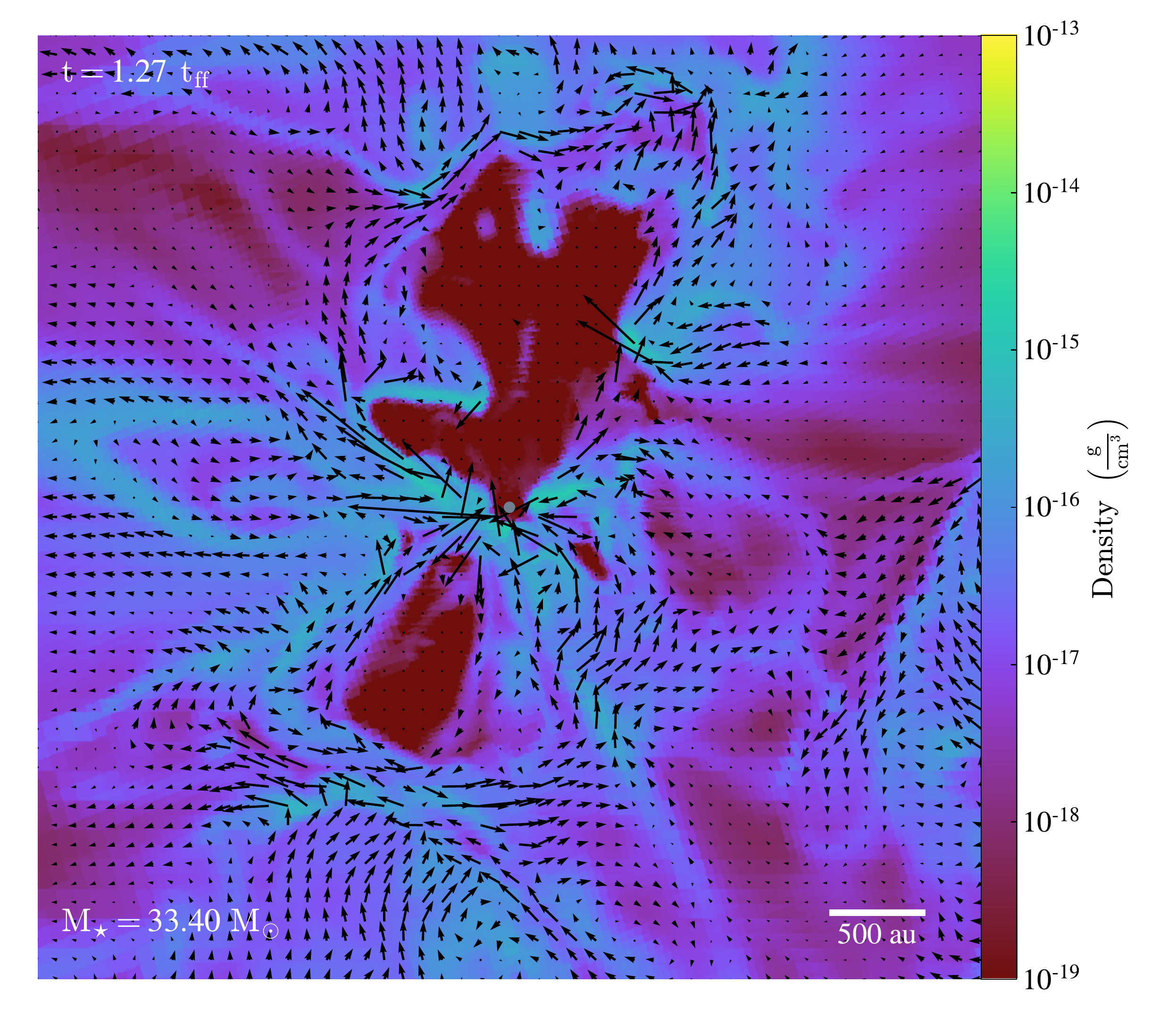}}%OFB_densityBubble_Bfield.pdf}} 
\caption{
\label{fig:Bquiver1}
Density slice for the second to last snapshot in run \hrtofb\ shown in Figure~\ref{fig:densityzoom} with magnetic field vectors over-plotted. The area shown is (5,000 au)$^2$. The massive star is located at the center of the panel, marked by the gray circle, and the slice is oriented so that the angular momentum axis of the gas within a radius of 250 au from the primary star points up.} %\mrknote{You should add a scale arrow to indicate what magnetic field strength a given arrow length corresponds to.}}
\end{figure}

We also find that radiative feedback is less important when magnetic fields are present because magnetic confinement inhibits the expansion of the radiation pressure dominated bubbles: as the cavities expand their dense shells sweep up magnetic flux, amplifying the field strength in the shells \citep{Krumholz2007c}. The increase in magnetic tension at the shells suppresses expansion, as shown in Figure~\ref{fig:Bquiver1}, which zooms in on the second to last panel of run \hrtofb\ in Figure~\ref{fig:densityzoom} and has the magnetic field vectors over-plotted to highlight the increased field strength in the bubble shells. The magnetic field structure also shows that the field lines are predominantly parallel to the shells, thereby opposing the radiation-pressure dominated bubble expansion.

\begin{figure*}
\centerline{\includegraphics[trim=0.2cm 1.5cm 0.2cm 1.25cm,clip,width=1\textwidth]{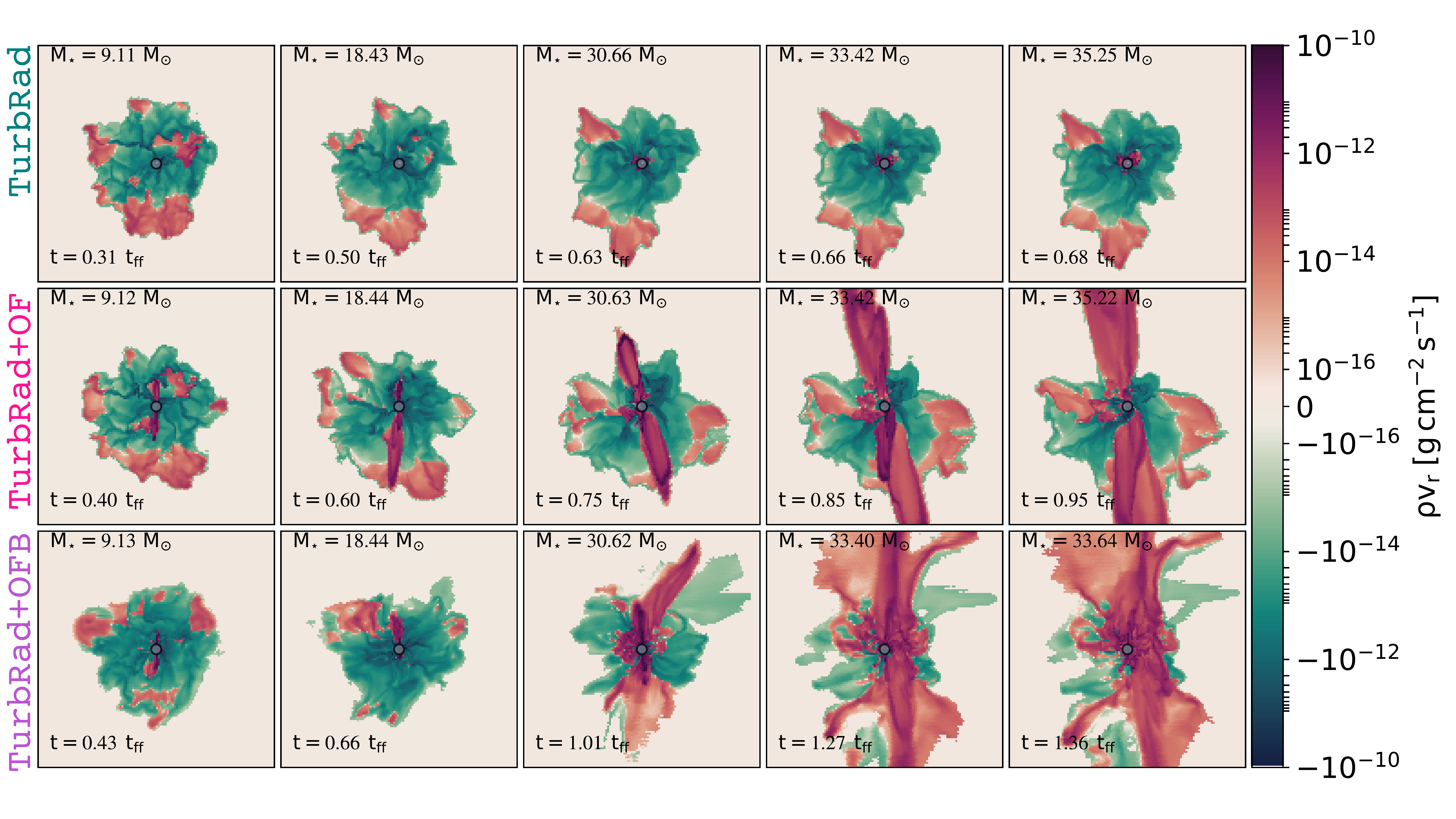}}%Turb_my_rhovr_Both_multipanel_d500AUbig_Jstar.pdf}}
\caption{
\label{fig:rhovr}
Slices of the gas radial momentum with respect to the primary star with a diverging color scale chosen to highlight material that is moving toward (negative values of $\rho v_{\rm r}$) and away from (positive values of $\rho v_{\rm r}$) the star for runs \hrt\ (top panels) and \hrtof\ (bottom panels) at different times. Each panel is (0.4 pc)$^2$ in area and the gray star at the center of each panel denotes the location of the primary massive star. The slices are oriented so that the angular momentum axis of the protostar points up to highlight the radial momentum from the outflows.
}
\end{figure*}

In addition to radiation pressure, runs \hrtof\ and \hrtofb\ show that feedback associated with collimated protostellar outflows, which are present throughout the star formation process (i.e., from low- to high-masses), drive high-velocity entrained outflows whose opening angle and extent increase with time. The outflows eventually break out of the core and eject material when the primary star reaches $ \sim 30 \, M_{\rm \odot}$. This behavior is more apparent in Figure~\ref{fig:rhovr}, which shows thin density projections of the radial momentum, $p_r = \rho v_r$, with respect to the primary star. Negative values of $p_r$ denote material that is falling towards the star whereas positive values denote material that is moving away from the star. Comparison of Figures~\ref{fig:densityzoom} and \ref{fig:rhovr} show that the entrained outflowing material due to feedback from protostellar outflows is not necessarily aligned with the low-density cavities associated with the expanding radiation pressure dominated bubbles. 

\subsection{Primary Protostar Properties}

\begin{figure*}
\centerline{\includegraphics[trim=0.2cm 0.2cm 0.2cm 0.2cm,clip,width=1\textwidth]{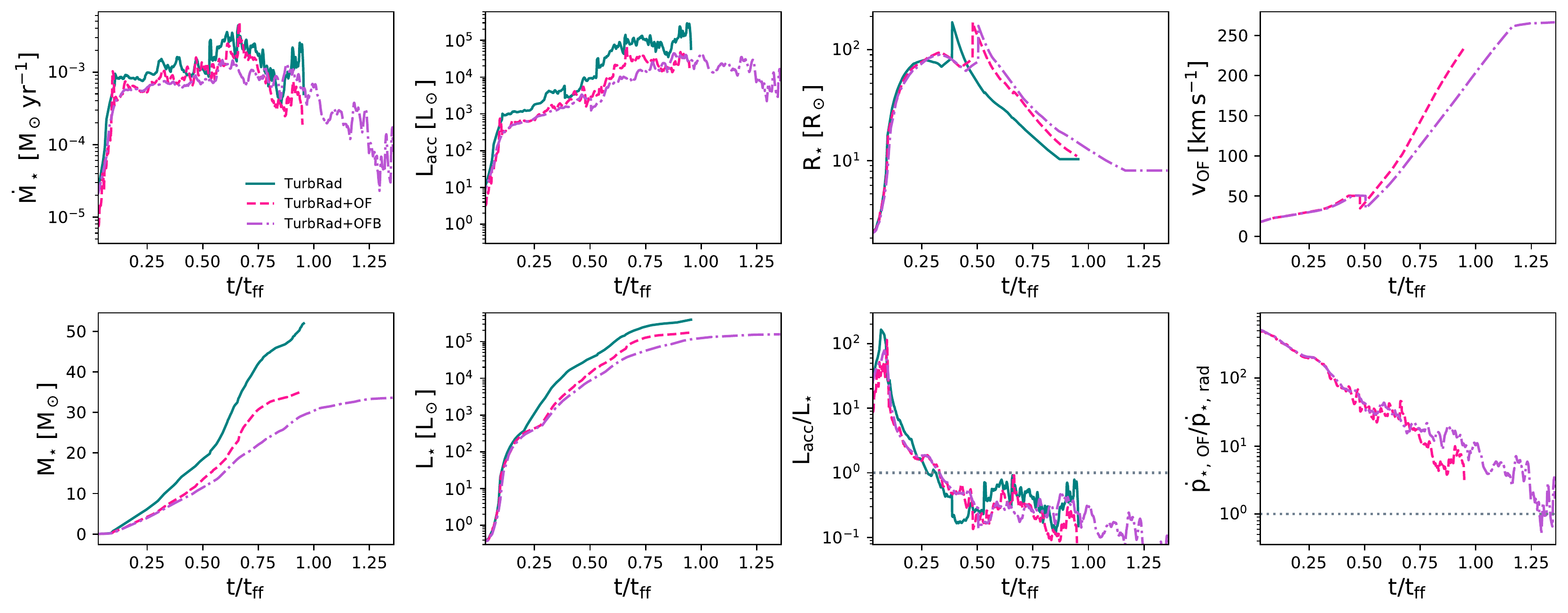}}%OFallRuns_starProps_tsim.pdf}}
\caption{
\label{fig:star_props}
Protostar properties as a function of simulation time for runs \hrt\ (teal solid lines), \hrtof\ (pink dashed lines), and \hrtofb\ (purple dot-dashed lines). The top row shows the primary star's accretion rate, accretion luminosity, radius, and outflow velocity. The bottom row shows the primary star's mass, internal (stellar) luminosity, ratio of the accretion to internal luminosity, and the ratio of the rate of momentum deposition for outflows with respect to direct radiation including contributions from both the internal and accretion luminosities. The gray dashed line in the two bottom right panels denote where these ratios are equal to 1.
}
\end{figure*}

We show the properties of the primary protostar as a function of simulation time for runs \hrt\ (teal solid lines), \hrtof\ (pink dashed lines), and \hrtofb\ (purple dot-dashed lines) in Figure~\ref{fig:star_props}. The far left column shows the accretion rate (top panel) and primary stellar mass (bottom panel), demonstrating that the mass growth rate is fastest for run \hrt\ since accreted material is not lost to outflows, as happens in runs \hrtof\ and \hrtofb. The growth rate is slowest for run \hrtofb\ because, in addition mass loss by outflows, magnetic pressure slows down the gravitational collapse of the core. However, the overall accretion history for all three runs are similar in shape. The one difference we see between the runs is that the accretion rate decreases at late times for run \hrtofb, which we run for the longest time, because feedback (aided by magnetic pressure) becomes sufficient to begin dispersing the core, thereby reducing the infall of material that can be accreted by the primary star. The accretion rate may also decrease because magnetic braking inhibits the formation of an optically thick accretion disk in run \hrtofb, a possibility that we discuss in more detail in Section~\ref{sec:disk}.

The accretion luminosity, stellar luminosity, and ratio of these two quantities are shown in the top middle left panel, bottom middle left panel, and bottom left panel, respectively. Early in all runs, when the star is less than $\sim$ several $M_{\rm \odot}$, the accretion luminosity is larger than the stellar luminosity due to the high accretion rates inherent to massive star formation. As the primary star increases in mass and contracts to the zero age main sequence (ZAMS), as shown in the radial evolution plot in the top middle right panel, the stellar luminosity becomes larger than the accretion luminosity. Once this transition occurs, the accretion luminosity varies between $\approx$10\%-100\% of the stellar luminosity, suggesting that the accretion luminosity is non-negligible in massive star formation until late times when the accretion rate tapers off.

We show the outflow velocity and ratio of the rate of momentum deposition from outflows, $\dot{p}_{\rm OF} = \dot{M}_{\rm OF} v_{\rm OF}$, to the rate of momentum deposition from direct radiation, $\dot{p}_{\rm rad} = (L_{\rm \star} + L_{\rm acc})/c$, in the top far right and bottom far right panels, respectively. The outflow velocity weakly increases until $t\sim 0.5 t_{\rm ff}$ because the primary protostar's radius and surface escape speed gradually increase. However, once the star begins to contract to the ZAMS, the outflow velocity increases rapidly. Throughout the majority of the accretion history the rate of momentum deposition by outflows is much larger than the rate of momentum deposition by radiation. At early times this ratio is of the order of $\sim$ few $\times 100$, and it eventually decreases to a factor of a few by the end of run \hrtof\ and a factor of $\sim 1$ by the end of run \hrtofb. Therefore, we find that the momentum input by outflows dominates over the momentum input by radiation throughout the majority of the star formation process.

\subsection{Primary Protostar Angular Momentum Evolution}
\label{sec:angmom}

\begin{figure}
\centerline{\includegraphics[trim=0.25cm 0.25cm 0.25cm 0.25cm,clip,width=1\columnwidth]{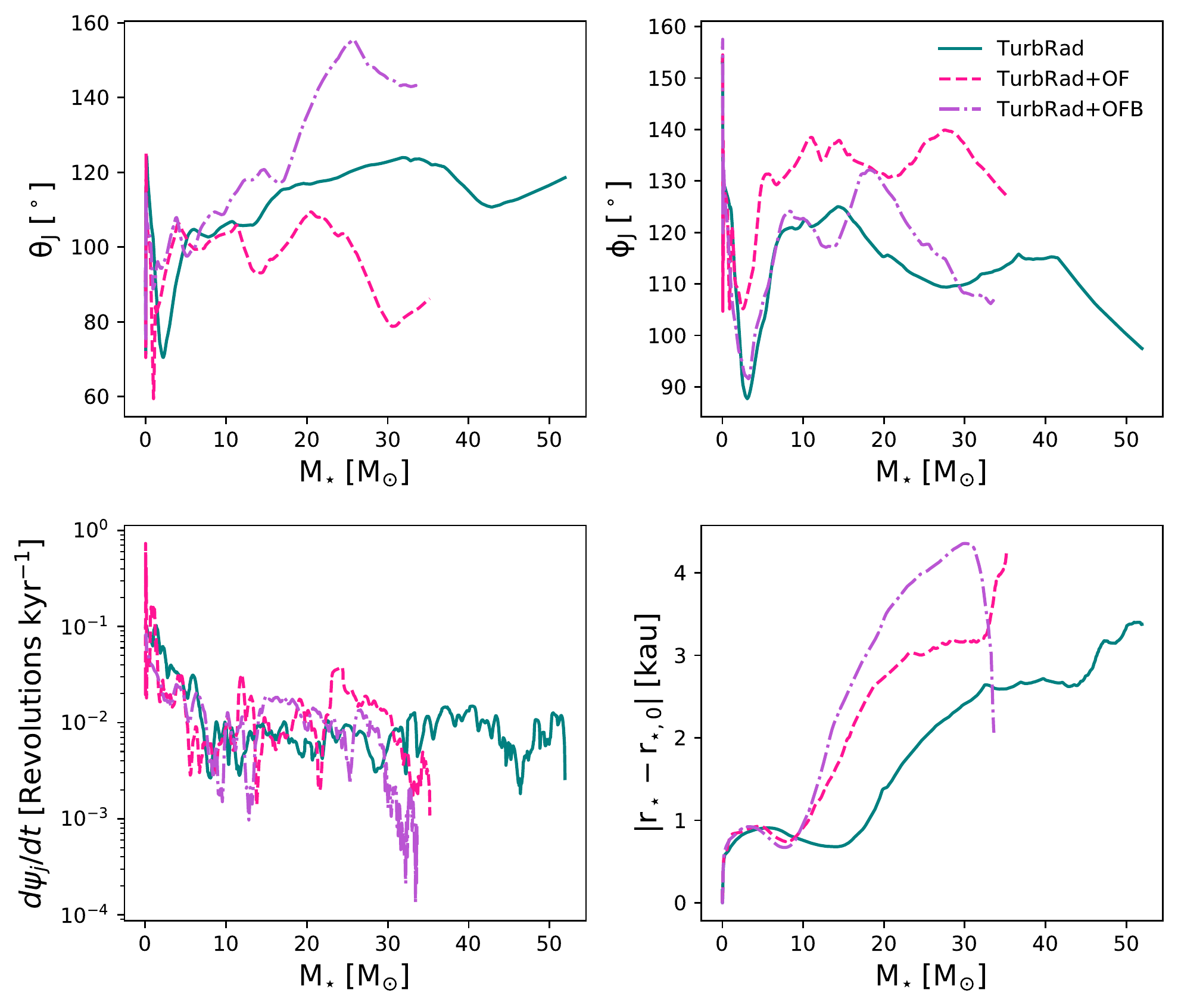}}%OFallRuns_AngMom_Mstar.pdf}}
\caption{
\label{fig:angmomm}
Primary star's position and angular momentum evolution as a function of the primary stellar mass for runs \hrt\ (solid teal lines), \hrtof\ (dashed pink lines), and \hrtofb\ (dot-dashed purple lines). The top row shows the angular momentum evolution for the primary star in spherical coordinates $\theta_j = \arctan{(j_y/j_x)}$ (left column) and $\phi_j = \arccos{(j_z/|\mathbf{j}|)}$ (right column) where $\mathbf{j} =(j_x,j_y, j_z)$ is the direction of the primary star's  spin axis in Cartesian coordinates. The bottom left panel shows the rate of change of the angle traced by the star's spin axis and the bottom right panel shows the massive star's position with respect to the location where it was formed within the core.
}
\end{figure}

We show the evolution of the primary star's position and spin axis in Figure~\ref{fig:angmomm} as a function of primary stellar mass, respectively. The top row of this figure shows the evolution of the primary star's spin axis in spherical coordinates: $\theta_j = \arctan{(\hat{j}_y/\hat{j}_x)}$ (left panel) and $\phi_j = \arccos{(\hat{j}_z)}$ (right panel) where $\mathbf{\hat{j}} =(\hat{j}_x,\hat{j}_y, \hat{j}_z)$ is the unit vector describing the direction of the primary star's spin axis in Cartesian coordinates. This figure demonstrates that the momentum and angular momentum accreted by the star cause the star to move, and lead the primary star's spin axis to precess, especially at early times when the star is low in mass (i.e., $M_{\rm \star} \lesssim 10 M_{\rm \odot}$). We describe the impact this has on the outflow structure in more detail in Section~\ref{sec:outflows}. 

The bottom left panel of Figure~\ref{fig:angmomm} show the rate of change of the angle traced by the primary star's spin axis defined as
\begin{equation}
    \frac{d\psi_j}{dt} = \frac{1}{2\pi} \left|\frac{d\hat{\mathbf{j}}}{dt}\right|
    %\frac{d\psi_j}{dt} = \frac{1}{\pi}\frac{d}{dt}\arccos{(\delta %\mathbf{\hat{j}} \cdot \delta \mathbf{\hat{j}})},
\end{equation}
where we have divided this quantity by $2\pi$ to convert from radians to revolutions where one revolution corresponds to the spin axis precessing by 360$^{\circ}$. This panel in Figure~\ref{fig:angmomm} shows that the precession decreases and $d\psi_j/dt$ flattens once the star has a mass $\gtrsim 10 \; M_{\rm \odot}$. Thereafter the rate of spin precession stays relatively constant and low until accretion begins to taper off, at which point precession stops as well.

\subsection{Accretion Disk Evolution}
\label{sec:disk}

\begin{figure*}
\centerline{\includegraphics[trim=0.2cm 4cm 0.2cm 3cm,clip,width=1\textwidth]{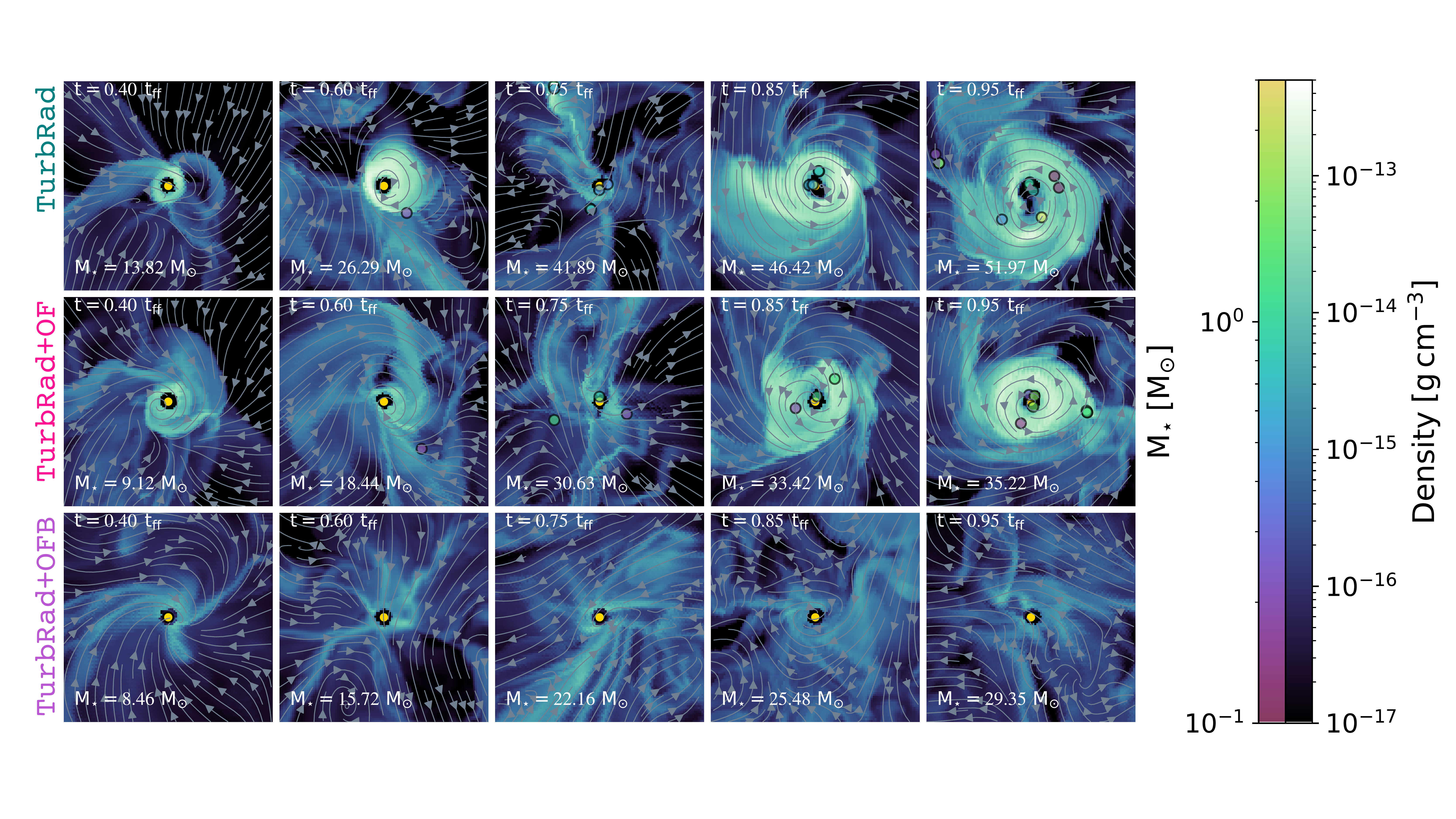}}%Turb_disk_All_multipanel_Slice_Time.pdf}}
\caption{
\label{fig:disk}
Density slices in runs \hrt\ (top row), \hrtof\ (middle row), \hrtofb\ (bottom row) shown at equal times; the slices have been oriented so that the angular momentum axis of the material within 250 au of the primary star points out of the page, in order to highlight the accretion disk. In each panel, the most massive star is at the center, and the region shown around it is (2000 au)$^2$ in size. Velocity streamlines and companion stars with masses greater than 0.04 $M_{\rm \odot}$ are over-plotted on all panels. The color of the star indicates its mass, as shown in the colorbar.} 
\end{figure*}

Figure~\ref{fig:disk} shows a series of density slices of the accretion disk that forms around the massive star with velocity streamlines over-plotted. Here, we compare each simulation at equal times since the accretion disk structure depends on the angular momentum content of the collapsing core. This figure shows that a noticeable high-density accretion disk (i.e., a resolved accretion disk with a radius larger than the 80 au accretion zone radius of the sink particle) forms around the primary stars in runs \hrt\ and \hrtof. However, we do not see a resolved accretion disk in run \hrtofb. 

This difference is almost certainly a result of magnetic braking. The accretion disk grows in size as runs \hrt\ and \hrtof\ progress owing to conservation of angular momentum: as time advances, the material reaching the vicinity of the primary star began its infall from greater distances, and thus has larger net angular momentum and therefore will be circularized at a distance farther from the star.  Magnetic fields suppress this effect by carrying angular momentum away from the infalling material and farther out into the surrounding core suppressing disk growth \citep{Seifried2011a, Commercon2011b, Myers2013a}. Non-ideal effects, \add{which remains an active field of study in protostellar disk formation \citep[e.g., see detailed reviews by][]{Wurster2018a, Zhao2020a},} may reduce the effect of magnetic braking leading to small accretion disks or toroids \add{or mitigate the magnetic braking catastrophe entirely  \citep{Kolligan2018a, Wurster2019a}}. However, we do not include these effects, and, even if they were included, it is still possible that any resulting disk would be unresolved in our simulation. 

\begin{figure} 
\centerline{\includegraphics[trim=0.25cm 0.25cm 0.25cm 0.25cm,clip, width=1\columnwidth]{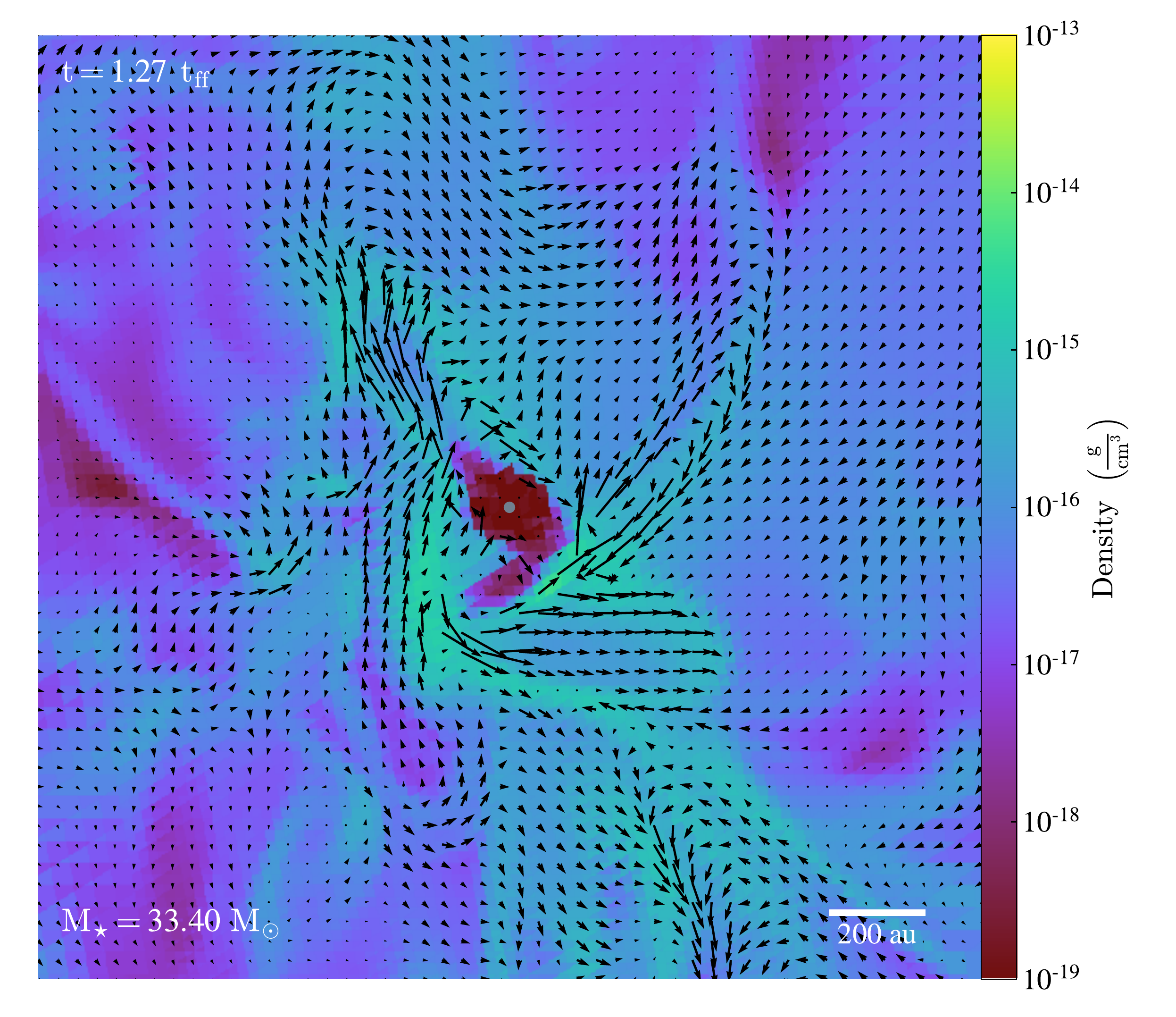}}%OFB_densityDisk_Bfield.pdf}} 
\caption{
\label{fig:Bquiver2}
Similar as the bottom right panel of \autoref{fig:disk} but for the snapshot shown in Figure~\ref{fig:Bquiver1}, and with magnetic field vectors rather than velocity streamlines over-plotted.
}
\end{figure}

We do note that at early times for run \hrtofb\ in Figure~\ref{fig:disk} a small disk-like structure forms around the massive star, but this disappears later in the simulation. Instead, radiation pressure near the star yields a magnetically confined low-density bubble surrounding the star that causes the accretion rate to drop at late times. We show this structure in Figure~\ref{fig:Bquiver2}, which shows the density structure near the primary star near the end of run \hrtofb\ with magnetic field vectors over-plotted.

\subsection{Outflow Properties}
\label{sec:outflows}

Thus far we have shown that outflows have a decisive effect: they deliver far more momentum to a protostellar core than radiation pressure,  they entrain a significant amount of mass, and they reduce the influence of radiation pressure by providing channels through which radiation can escape. In this section we explore the structure and properties of the outflows in more detail.

\begin{figure*}
\centerline{\includegraphics[trim=0.2cm 7.25cm 0.2cm 7.0cm,clip,width=1\textwidth]{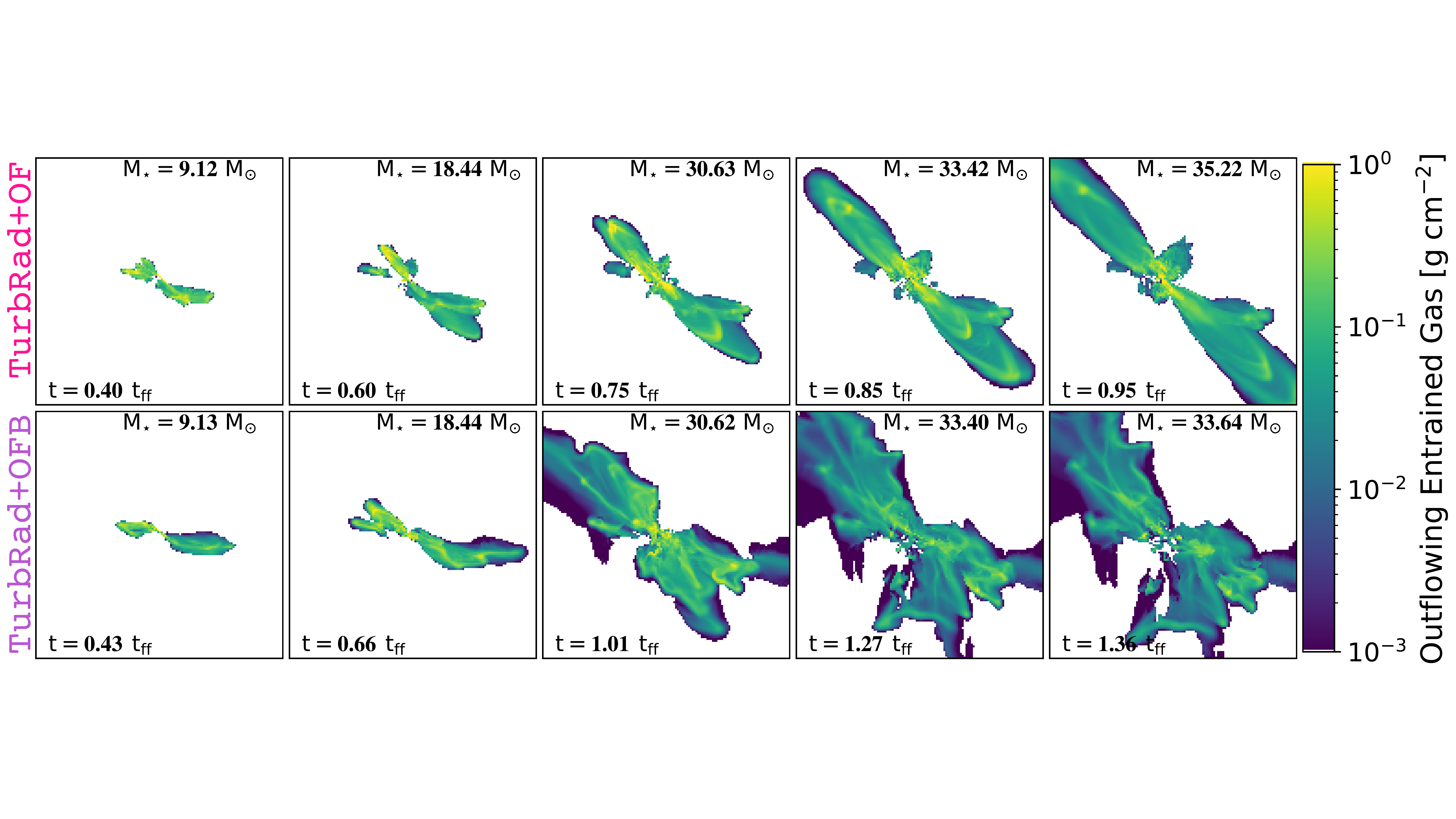}}%Turb_massEntrainedOut_Both_multipanel_OFB.pdf}}
\caption{
\label{fig:outflows}
Projections of the density of the entrained outflows along the $yz$ plane that are moving away from the primary star ($v_{\rm r} > 0$) for runs \hrtof\ (top row) and \hrtofb\ (bottom row).
}
\end{figure*}

\subsubsection{Density Structure}

We show projections of the material entrained by outflows for run \hrtof\ (top row) and run \hrtofb\ (bottom row) in Figure~\ref{fig:outflows}. We define entrained material as consisting of all cells whose mass contains at least 5\% of the launched material (i.e., cells where $f_t=\rho_{\rm OF}/\rho \ge 0.05$); recall that we add a passively advected scalar to the outflow material we inject, which allows us to measure $\rho_{\rm OF}$ precisely for each cell. We only include gas that has a positive radial velocity, $v_r>0$, with respect to the primary star and subtract the launched material from the total density so that we only include contributions of entrained material. We note that our definition does include contributions from the low-mass companion stars as well as the primary star since we are not able to trace the ejected quantities for individual stars. However, this should only be a minor effect since most of the injected outflow mass is from the primary star.

Figure~\ref{fig:outflows} shows that the outflow structure is not steady. At early times there appear to be multiple outflows present. In reality the primary star dominates outflow production at all times, but, as discussed in Section~\ref{sec:angmom}, the star's angular momentum axis changes rapidly when its mass is low, causing outflows to be launched in multiple directions. The precession eventually decreases as the star increases in mass leading to a steadier outflow. Eventually the opening angles of the entrained outflows broaden such that the multiple outflows from the massive star merge. This effect leads to wider outflows as shown in the middle column of Figure~\ref{fig:outflows}; the outflow broadening is much larger in run \hrtofb\ as compared to run \hrtof. Hence, magnetic fields produce lower density and broader, less-collimated outflows.

\begin{figure}
\centerline{\includegraphics[trim=0.2cm 0.2cm 0.2cm 0.2cm,clip,width=0.75\columnwidth]{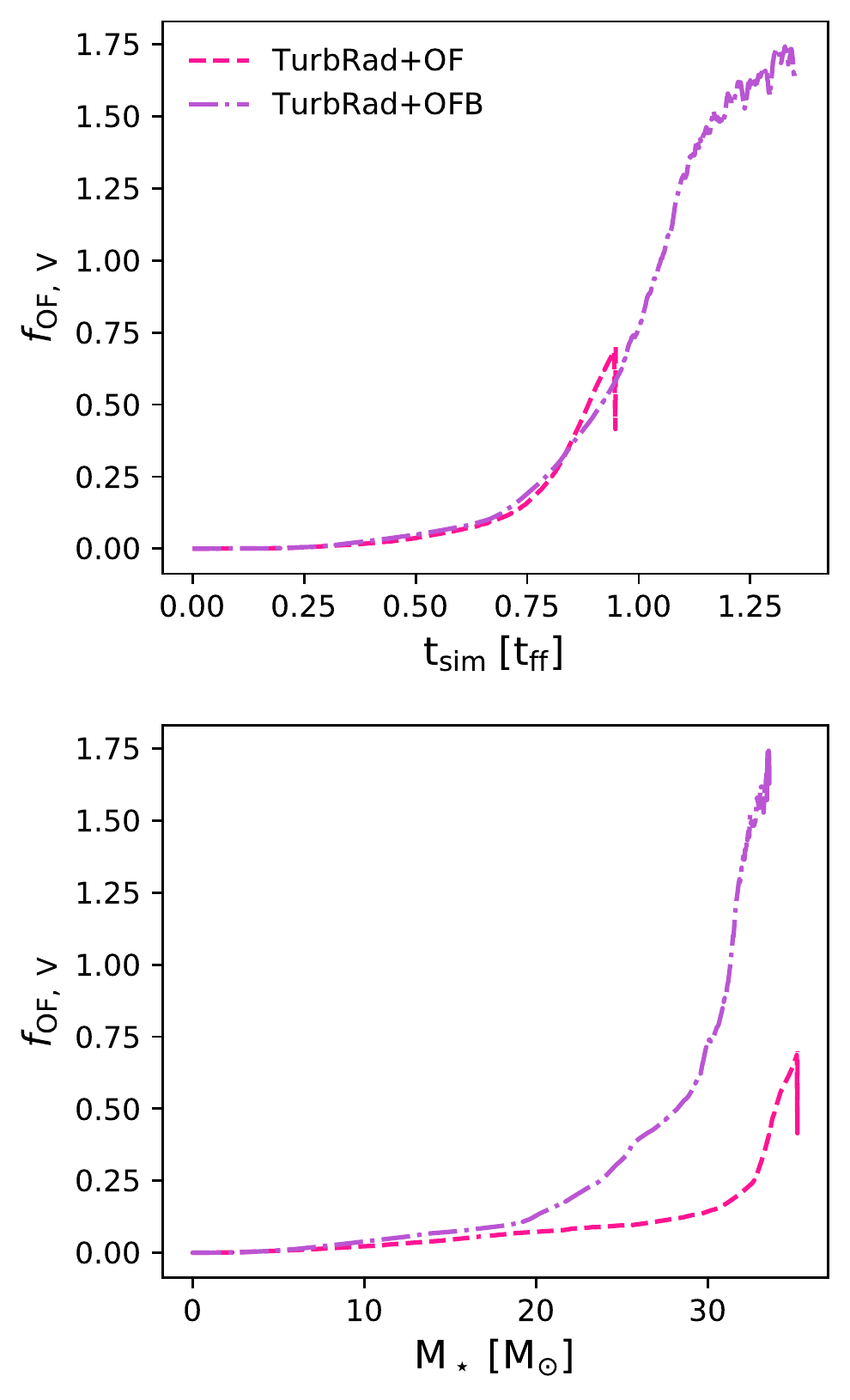}}%OF_volProps.pdf}}
\caption{
\label{fig:ofvol}
Outflow volume fraction as a function of simulation time (top panel) and primary stellar mass (bottom panel) for runs \hrtof\ (pink dashed lines) and \hrtofb\ (purple dot-dashed lines).} 
\end{figure}

We quantify the outflow broadening in Figure~\ref{fig:ofvol}, which shows the volume filling fraction of the outflows and entrained gas for runs \hrtof\ (pink dashed lines) and \hrtofb\ (purple dot-dashed lines) as a function of time (top panel) and primary stellar mass (bottom panel). We calculate this value by summing over the volume of all cells whose mass contain launched material normalized to the initial core volume: $f_{\rm OF, \, V} = \sum_i dV_{\mathrm{OF}, i}/V_{\rm core,\, init}$ where $V_{\rm core,\, init} = \frac{4}{3} \pi R^3_{\rm c} $. We find that, as a function of simulation time, the volume filling fractions are roughly the same for both simulations and they slowly increase up to $t\approx 0.8 \; t_{\rm ff}$. After this time, the entrained volume fraction increases much more rapidly for both runs. The rapid increase corresponds to the point where outflows begin to break out of the the initial core and expand freely into the low-density medium outside it. However, it is interesting to notice that, despite their similarity in time, there is a significant difference between the runs when we study the behavior in terms of primary star mass: the upturn in the volume filling fraction occurs when the star reaches $\sim 20 M_{\rm \odot}$ in run \hrtofb\ as compared to $\sim 30 M_{\rm \odot}$ in run \hrtof. This suggests that less momentum is required for outflow breakout in run \hrtofb, likely because magnetic levitation reduces gravitational confinement of the core \citep{Shu2004a}.

\subsubsection{Mass, Momentum, and Energy Budgets}
\label{sec:ofbudgets}

\begin{figure}
\centerline{\includegraphics[trim=0.25cm 0.25cm 0.25cm 0.25cm,clip,width=0.85\columnwidth]{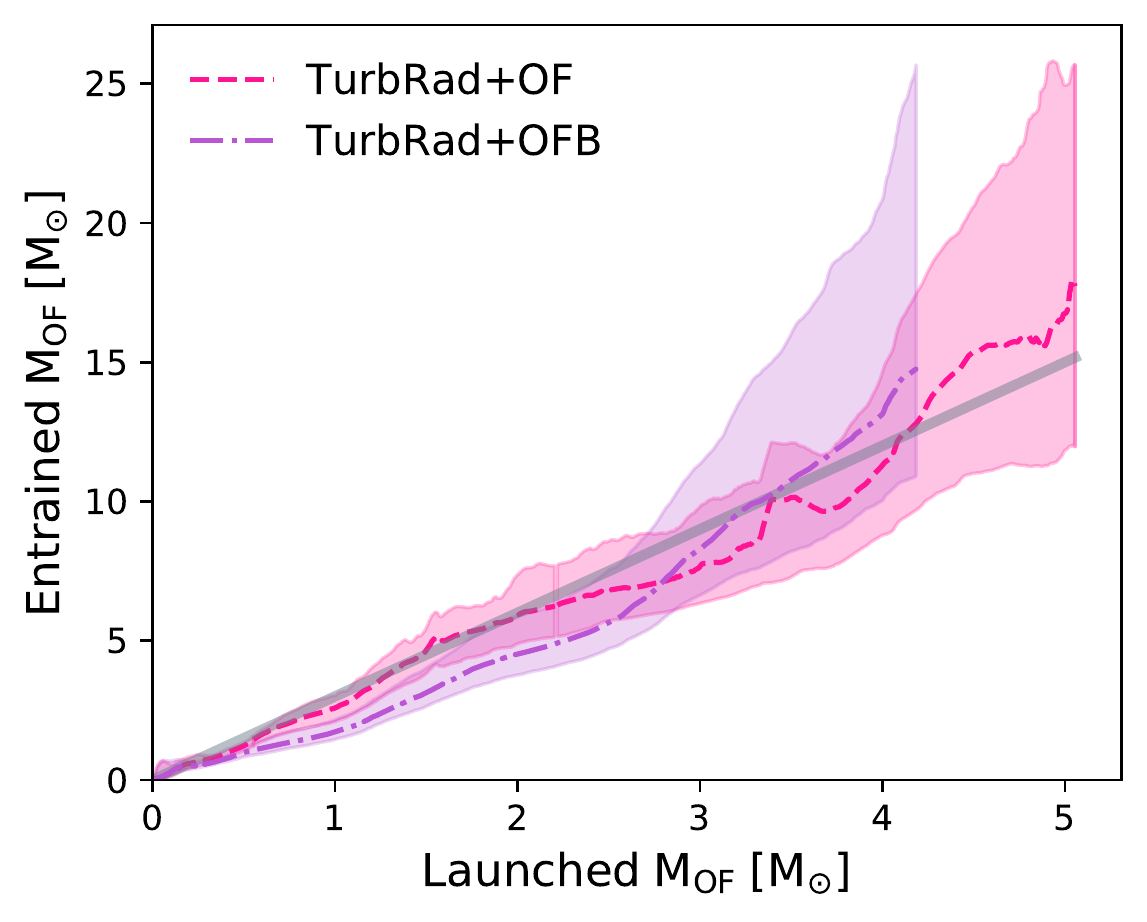}}%OF_outflowBudgets.pdf}}
\caption{
\label{fig:ofprops}
Entrained outflow versus injected mass for runs \hrtof\ (pink dashed lines) and \hrtofb\ (purple dot-dashed lines). \add{The shaded pink and purple regions denote the entrained outflow mass where we consider cells whose density contain between 1-10\% of outflow material for runs \hrtof\ and run \hrtofb, respectively.} The gray line has a slope of three (i.e., $M_{\rm OF, \, ent} = 3 M_{\rm OF, \, inj}$) .
}
\end{figure}

\begin{figure}
\centerline{\includegraphics[trim=0.25cm 0.25cm 0.225cm 0.25cm,clip,width=0.85\columnwidth]{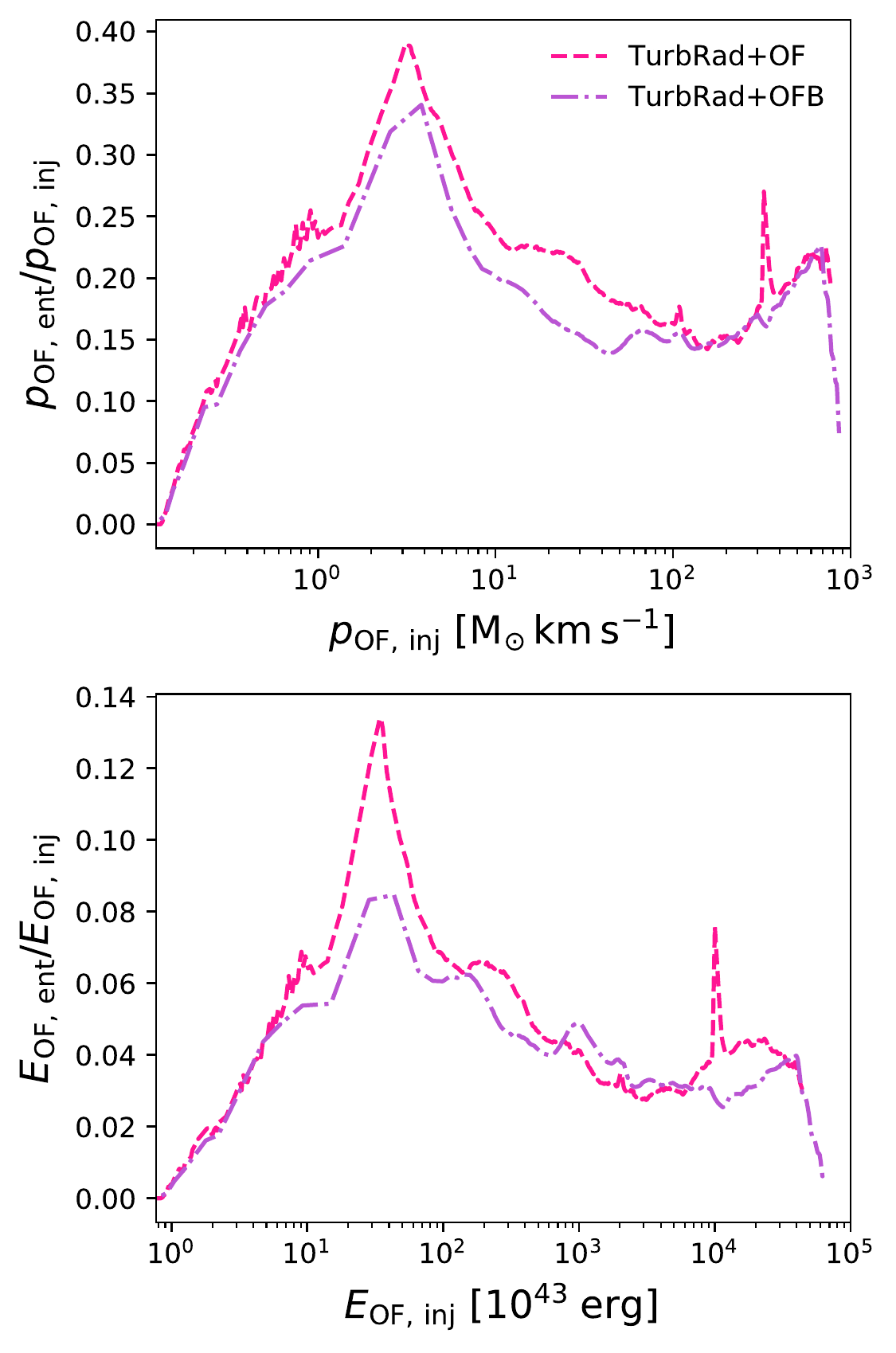}}%outflow_boost_plot.pdf}}
\caption{
\label{fig:ofprops2}
Ratio of the entrained outflow to injected momentum (top panel) and kinetic energy (bottom panel) as a function of the injected momentum and kinetic energy, respectively, for runs \hrtof\ (pink dashed lines) and \hrtofb\ (purple dot-dashed lines). 
}
\end{figure}

We show the launched outflow mass versus the entrained outflow mass in Figure~\ref{fig:ofprops} for runs \hrt\ (pink dashed line) and \hrtofb\ (purple dashed line)\add{, in which we consider that material is entrained if the cell contains at least 5\% of the launched outflow material. The shaded regions denote the spread in the total entrained outflow mass considering cells that contain 1-10\% of the launched outflow material.} The over-plotted gray solid line has a slope of 3 showing that the entrained outflow mass is a factor of $\sim 2-3$ larger than the launched outflows mass. Hence, we find that feedback from collimated outflows and radiation pressure have a mass-loading factor of $\sim 2-3$ in massive star formation, so the total amount of mass directly removed is $\sim 50\%$ of the core (since the outflow at launch contains $\sim 20\%$ of the accreted mass). Additionally, Figure~\ref{fig:ofprops} shows that the entrained mass is quantitatively similar regardless if the core is magnetized or not. \add{The shaded regions in this Figure show that the largest variations in total entrained mass occur near the end of the simulation when we consider a low (e.g., 1\%) fraction of outflow to entrained material, likely due to advection of the launched material. However, a larger fraction (e.g., 10\%) yields quantitatively similar results of the total mass entrained as compared to when we consider cells that contain $\ge5\%$ of launched material.}

We also compare the entrained outflow momentum and kinetic energy to the injected outflow momentum and kinetic energy in Figure~\ref{fig:ofprops2}. We define the entrained outflow momentum and kinetic energy as 
\begin{equation}
p_{\rm ent} = \sum_i \rho_{\mathrm{ent},\, i} v_{i} dV_{i}
\end{equation}
\noindent
and
\begin{equation}
E_{\rm ent} = \sum_i  \frac{1}{2} \rho_{\mathrm{ent},\, i} v^2_{i} dV_{i}, 
\end{equation}
\noindent
respectively, where $\rho_{\mathrm{ent}, \, i}$ is the entrained outflow mass density, $dV_{i}$ is the cell volume, and $v_{i}$ is the gas velocity magnitude of cell $i$. Again, we remind the reader that we count a cell as entrained if it is at least 5\% outflow material by mass, and if the radial velocity, with respect to the primary star, is moving away from the primary star. Also note that these outflow quantities include contributions from both the primary and companion stars. However, this should make little difference since the injected and entrained outflow momentum and energy is dominated by the primary star regardless. For comparison, we compute the time-integrated injected outflow  momentum and kinetic energy from the stellar properties (see Figure~\ref{fig:star_props})  up to time $t$ as $p_{\rm OF,\, inj}(t) = \int_0^t \dot{M}_{\rm OF} v_{\rm OF} \, dt$ and $E_{\rm OF,\, inj}(t) = \int_0^t \frac{1}{2} \dot{M}_{\rm OF} v^2_{\rm OF} \, dt$, respectively.

We find that the entrained material contains $\sim25\%$ of the injected momentum and $\sim5\%$ of the injected kinetic energy. At late times, we also find that the momentum and kinetic energy contained within the entrained outflows are quantitatively independent of the core's magnetic field strength. The reduction in the entrained outflow momentum compared to that which was originally injected is likely due to mixing between the outflow material and ambient gas that has a negative radial momentum as a result of gravitational infall. Similarly, the reduction in energy is due to inelastic entrainment of material coupled with fast radiative cooling, since the cooling time is always short compared to any mechanical timescale. The fact that the reduction in energy is larger than the reduction in momentum suggests that outflows should be considered a momentum-driven rather than an energy-driven feedback mechanism.

%Comparing the launched versus entrained quantities allow us to constrain the mass-loading and momentum- and energy-boosts associated with protostellar outflows. In particular, we find that outflows have a mass loading factor of $\sim 2-3$, so the total amount of mass directly removed is $\sim 50\%$ of the core (since the outflow at launch contains $\sim 20\%$ of the accreted mass); the momentum carried is $\sim 2-3$ times the injected momentum, which indicates a modest amount of enhancement by radiation pressure and/or magnetic pressure not associated with the outflow; the energy loading factor is similarly relatively modest, which is not surprising since outflows are slow enough that the cooling time is always short. We find that these quantities are similar for both the magnetic and non-magnetic cores.

\subsection{Fragmentation and Companion Protostar Properties}
\label{sec:compstars}

\begin{figure*}
\centerline{\includegraphics[trim=0.25cm 0.25cm 0.25cm 0.25cm,clip,width=0.75\textwidth]{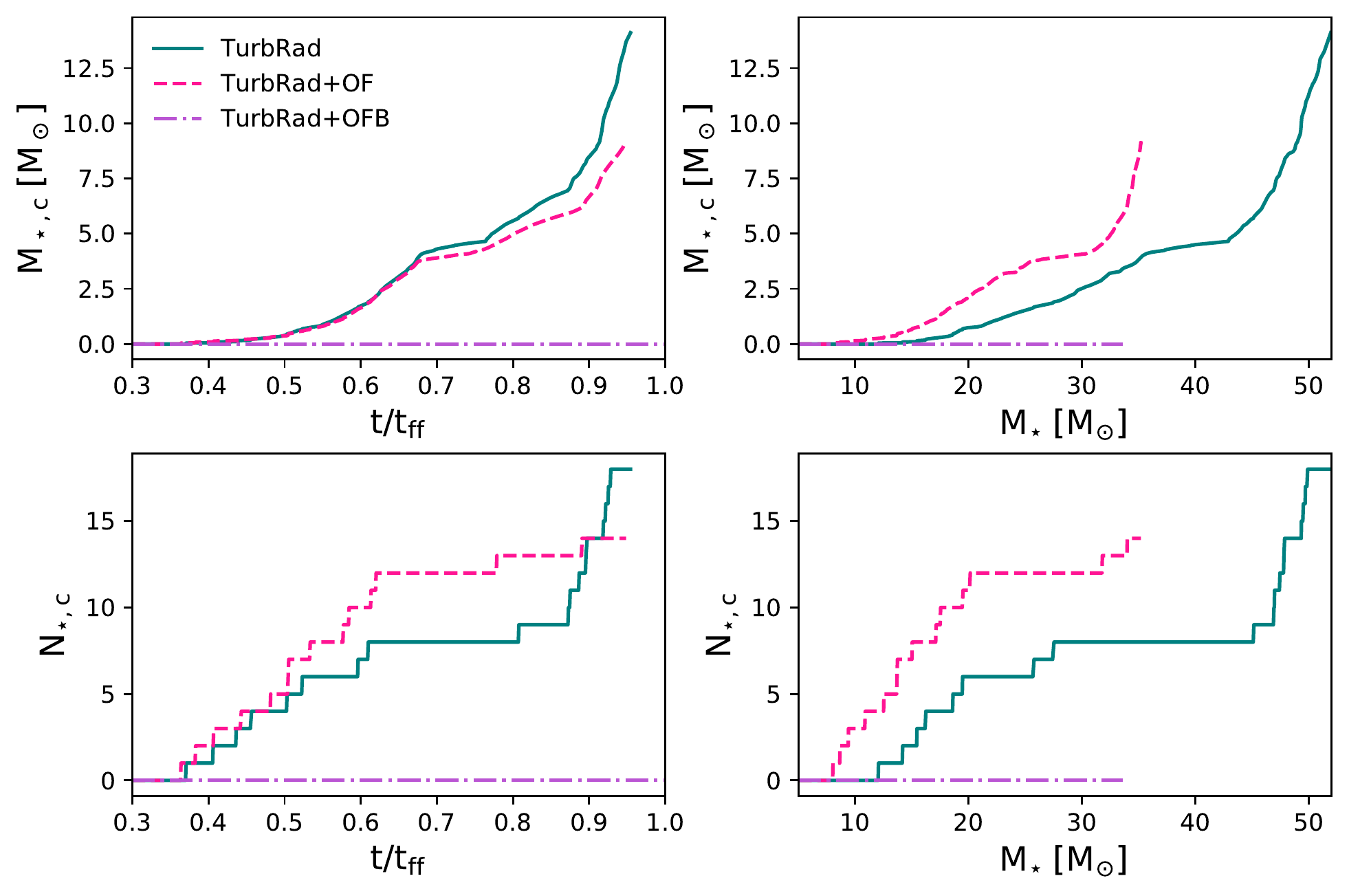}}%OFallRuns_CompStarProps.pdf}}
\caption{
\label{fig:compstars}
Total companion stellar mass (top row) and number of companion stars with $M_{\rm \star} > 0.04 M_{\rm \odot}$ (bottom row) in runs \hrt\ (teal solid lines) and run \hrtof\ (pink dashed lines) as a function of simulation time (left column) and primary stellar mass (right column). 
}
\end{figure*}

Companion stars form via turbulent fragmentation, in which over-densities in the surrounding core envelope collapse to form stars; and via disk fragmentation, in which the accretion disk becomes gravitationally unstable and fragments \citep{Kratter2006a, Rosen2019a}. We show the total companion stellar mass (top row) and number of companion stars (bottom row) as a function of simulation time (left column) and stellar mass (right column) for runs \hrt\ (solid teal lines), \hrtof\ (dashed pink lines), and \hrtofb\ (dot dashed purple lines) in Figure~\ref{fig:compstars}. This Figure shows that run \hrtofb\ does not form any companion stars because magnetic pressure suppresses turbulent fragmentation and additionally a resolved gravitationally unstable accretion disk does not form around the primary star \citep[e.g.,][]{Commercon2011b}.  

We find that from $t \approx 0.35-0.6 \; t_{\rm ff}$ companion stars form via turbulent fragmentation and at late times they form via disk fragmentation when the star is very massive for runs \hrt\ and \hrtof\ \citep{Rosen2019a}. For run \hrtof\ more companion stars form via turbulent fragmentation as compared to run \hrt\ because the primary star is less luminous due to its slower growth, thereby reducing radiative heating of the core material. In addition, outflows allow venting of the radiation field making radiative heating less effective. At late times we see that run \hrt\ forms more companion stars via disk fragmentation because the star is much more massive than the primary star in run \hrtof\ and becomes gravitationally unstable (e.g., see Figure~10 from \citet{Rosen2019a}). Given that we do not form companion stars via turbulent fragmentation in run~\hrtofb, but we form many in run~\hrtof\ we expect that weaker magnetic fields (i.e., cores with $\mu_{\rm \Phi} > 2$) should lead to a weaker degree of turbulent fragmentation \citep[e.g.,][]{Palau2013a, Myers2013a, Fontani2018a}.

\section{Discussion}
\label{sec:disc}
The purpose of this work is to understand how the interplay between magnetic fields and feedback from radiation pressure and collimated outflows effect the formation of massive stellar systems and affects the resulting entrained outflow structure from massive stars. Most notably, we find that momentum feedback from collimated outflows dominates over radiation pressure in massive star formation and ejects a significant fraction of molecular material from the core leading to low star formation efficiencies (SFEs; $\epsilon = M_{\rm \star, \; tot}/M_{\rm c}$ where $M_{\rm \star, \; tot}$ is the total stellar mass and $M_{\rm c}$ is the initial core mass). Additionally, the presence of magnetic fields slows the growth rate of massive stars, reduces core fragmentation, and leads to broader and lower density outflows as compared to the outflows that emanate from unmagnetized cores. 

In what follows, we discuss how magnetic fields and feedback from outflows and radiation pressure affects the growth rate of massive stars and the overall SFEs of massive cores in Section~\ref{sec:sfe}. Next, we discuss how feedback from outflows and radiation pressure affect the energetics of massive star forming cores in Section~\ref{sec:energetics}\add{.} %and present a new subgrid model based on our results in Section~\ref{sec:subgrid} to be used in star cluster and galaxy scale simulations that are unable to resolve the formation of individual stars.

\subsection{Influence of feedback and magnetic fields on the SFE of massive prestellar cores}
\label{sec:sfe}
\begin{figure}
\centerline{\includegraphics[trim=0.2cm 0.0cm 0.2cm 0.2cm,clip,width=0.75\columnwidth]{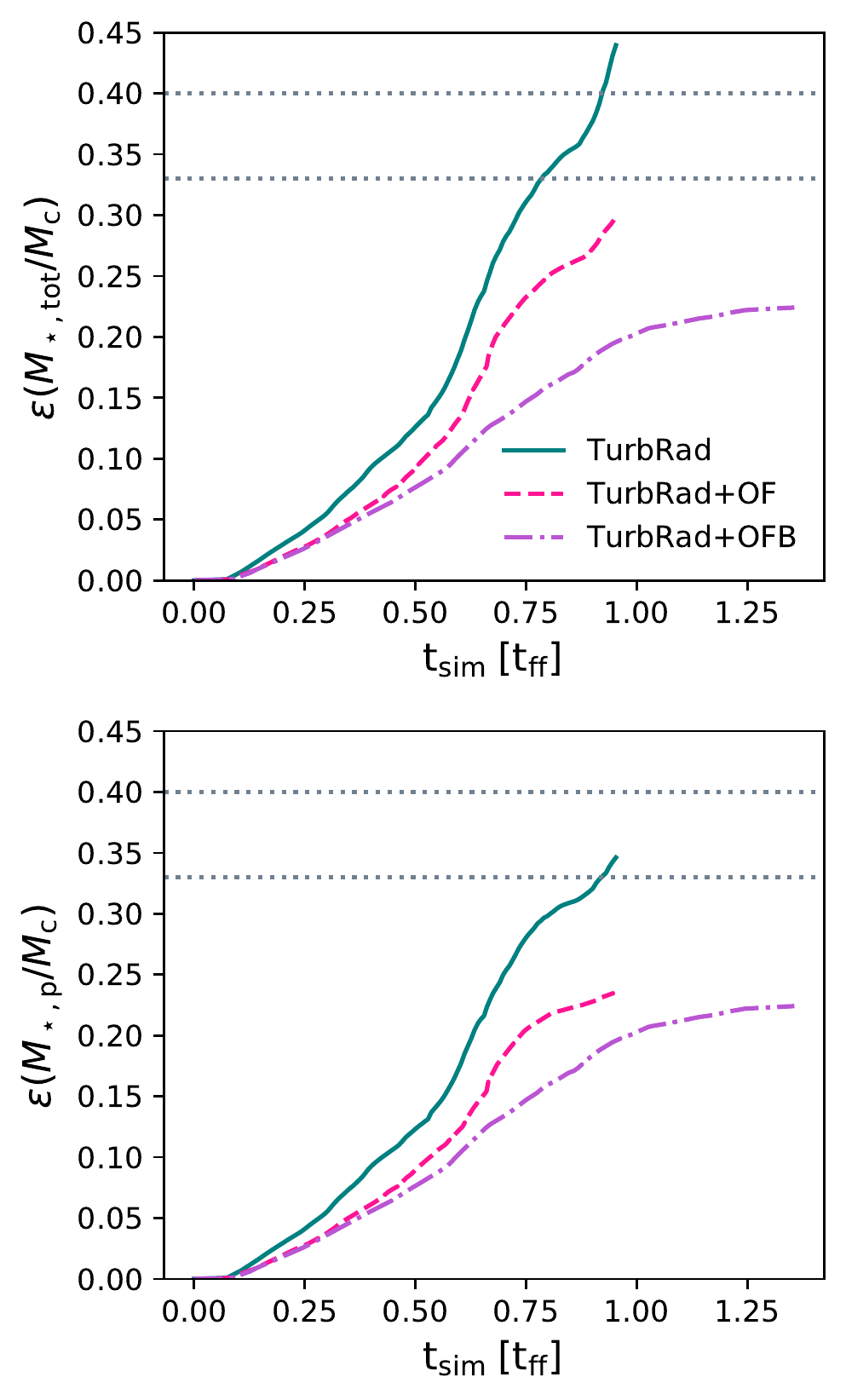}}%OF_SFE.pdf}}
\caption{
\label{fig:sfe}
Star formation efficiency, $\epsilon = M_{\rm \star, \, tot}\add{/}M_{\rm core}$ (top panel) as a function of simulation time for runs \hrt (solid teal line), \hrtof\ (pink dashed line), and \hrtofb\ (purple dash-dot line). The bottom panel shows the SFE of the core considering only the primary star. The gray dotted line in both panels denotes where $\epsilon = 0.33$ and $\epsilon = 0.4$ as determined by the core mass function for low-mass star formation by \citet{Alves2007a} and the SFE estimate for high-mass protocluster formation from \citet{Maud2015a}, respectively.
}
\end{figure}

Star formation is an inefficient process, likely as a result of stellar feedback, turbulence, and magnetic fields \citep[e.g.,][]{Matzner2000a, Offner2014a, Louvet2014a}. Observations of the stellar initial mass function (IMF) and prestellar core mass function (CMF) in low-mass star formation studies find a direct mapping between these two quantities in which the CMF is similar in shape to the IMF but offset by a factor $\sim 3$; yielding a core SFE of $\sim 33\%$. This finding suggests that magnetic fields and feedback from outflows may be responsible for the mass offset since radiation pressure is unimportant in low-mass star formation \citep{Alves2007a, Offner2017a}. 

Given the rarity of and large distances to massive star forming regions, and the fact that low-mass star formation likely occurs coevally with high-mass star formation in massive cores, it remains uncertain what SFEs are expected for massive star formation and which physical properties set the SFE of massive star forming cores \citep[e.g.,][]{Motte2018a, Pillai2019a}. \citet{Maud2015a} studied the core properties and SFEs of a sample of 89 distance limited ($d \lesssim$ 6 kpc) cores that hosted massive young stellar objects (MYSOs) and compact \hii\ regions. They found that the mass-luminosity plane of these sources is consistent with the luminosity expected from a proto-cluster that hosts at least one high-mass source and forms with a $\sim$40\% SFE, slightly larger than the value inferred from low-mass studies. 

Here, we use our simulations to determine the role that radiative and outflow feedback play in setting the SFE of magnetized and unmagnetized massive star forming cores and compare them to  the observed values described above. We show the total SFE for runs \hrt\ (teal solid line), \hrtof\ (pink dashed line), and \hrtofb\ (purple dot dashed line) in the top panel of Figure~\ref{fig:sfe} and the SFE considering only the primary star (i.e., $\epsilon  =  M_{\rm \star, \; p}/M_{\rm c}$)  in the bottom panel. Comparison of these two SFEs (total versus primary) allow us to determine how magnetic fields and feedback from radiation pressure and/or outflows affects the mass growth of the primary star and subsequent growth of the companion stars. We note that the SFE plotted for run \hrtofb\ in Figure~\ref{fig:sfe} is identical in both panels since this run does not form companion stars throughout the simulation run time. 

The top panel in Figure~\ref{fig:sfe} show that the SFE is largest for run \hrt, reaching $\sim 0.45$ at $t=0.95 \; t_{\rm ff}$; this is larger than the $\epsilon \approx 0.33$ and $\epsilon \approx 0.4$ (marked by the dashed gray lines) expected for low mass star formation \citep{Alves2007a} and high-mass protocluster formation \citep{Maud2015a}, respectively. This high SFE occurs because accreted mass is not lost to outflows, while radiation pressure alone is not sufficient at ejecting material from the core, at least for the simulation run time considered here. Additionally, the SFE for run \hrt\ continues to increase rapidly at late times suggesting it should further increase. These findings suggest that radiation pressure alone is not responsible for the SFEs observed in massive star forming environments.

Next, we find that the total SFE in the runs containing outflows are significantly lower than the SFE in run \hrt\ but that run \hrtof\ is larger than run \hrtofb, reaching a value of $\sim 0.3$ as compared to $\sim 0.2$ for run \hrtofb\ at $t=0.95 \; t_{\rm ff}$. The higher SFE for the non-magnetic core is attributed to the increase of stellar mass for the companion stars that are still accreting from core material at the end of run \hrtof\ and the slightly higher growth rate of the primary star. Hence, we find that outflows are required to set the SFEs expected for massive star formation and that the presence of strong magnetic fields leads to even lower SFEs. However, weaker magnetic fields and/or non-ideal MHD effects such as Ohmic resistivity, ambipolar diffusion, and the Hall effect, which we do not include in this work could lead to a higher degree of fragmentation and therefore higher SFEs in magnetized massive cores as seen in the numerical study of \citet{Fontani2018a} and the observational study of \citet{Palau2013a}.
The importance of these non-ideal effects remains substantially uncertain. Radiative effects should increase the ionisation fraction near massive protostars, and thus reduce the importance of non-ideal effects. If non-ideal effects do allow greater fragmentation than we find in run \hrtofb, this would likely lead to increase SFEs, perhaps close to those found by \citet{Maud2015a}.

When we only take into account the primary mass for calculating the SFE, as shown in the bottom panel, we find the SFE continues to increase at an accelerating rate by the end of run \hrt\ but tapers off at the end of run \hrtof\ and \hrtofb, suggesting that outflows, rather than radiation pressure, are required to shut off accretion onto the massive star at late times. Comparison with the top panel, which shows that the SFE for the total system in runs \hrt\ and \hrtof\ continues to increase when these simulations end, demonstrates that accretion onto the low mass companion stars leads to increased SFEs even when feedback from outflows cuts off the accretion flow onto the primary star. Our result agrees with the observational study presented in \citet{Pillai2019a}, which studied the protostellar content via outflow signatures in two infrared dark clouds, that found that low-mass protostars likely form co-evally at the earliest phase of high-mass star formation (i.e., $t \lesssim 50,000$ yrs). 

\subsection{Outflow Budgets}
\label{sec:energetics}

\begin{figure*}
\centerline{\includegraphics[trim=0.25cm 0.25cm 0.225cm 0.25cm,clip,width=0.75\textwidth]{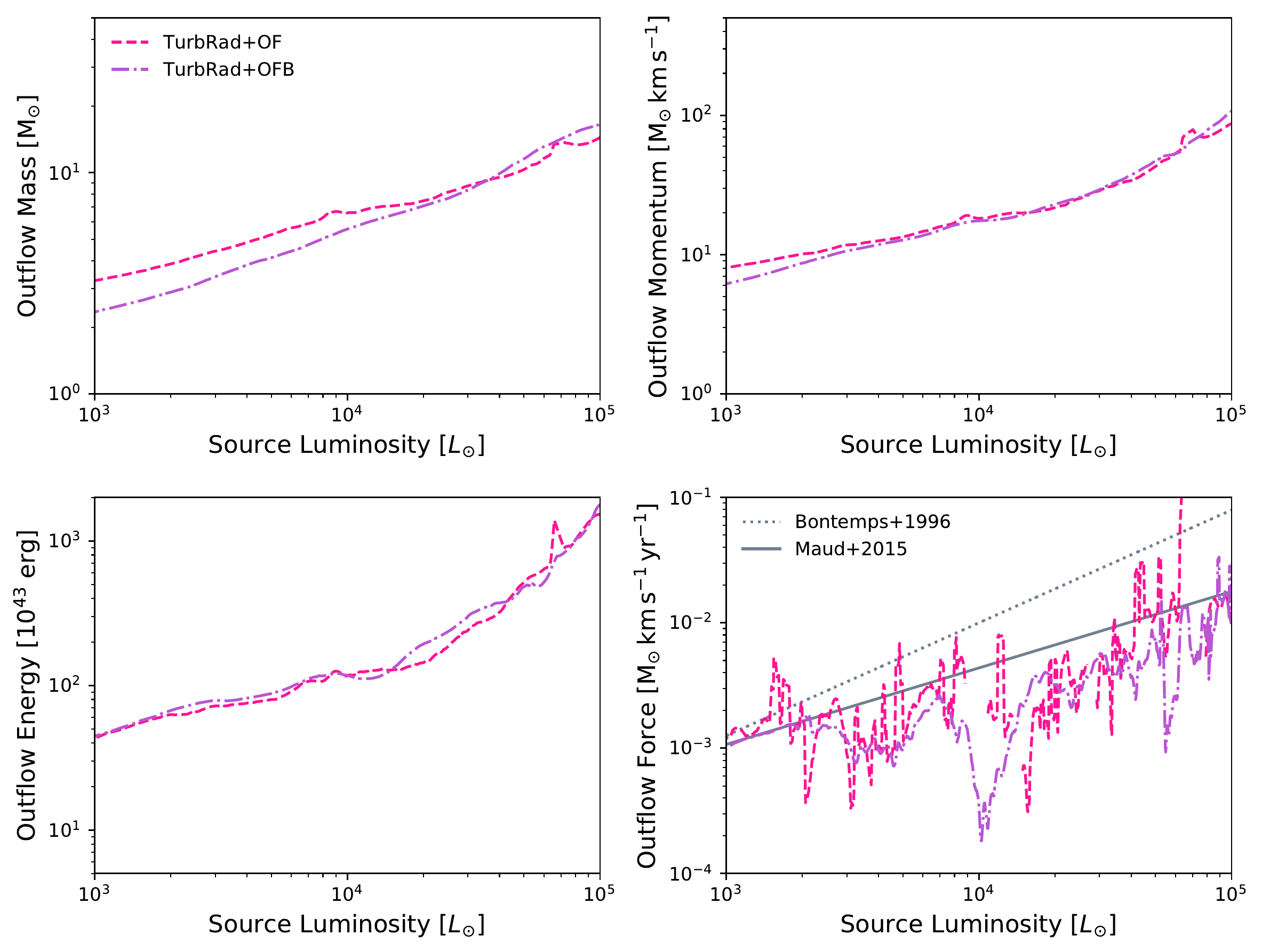}}%OFprops_Lstar.pdf}}
\caption{
\label{fig:OFpropsL}
Comparison of the outflow mass (top left panel), outflow momentum (top right panel), outflow energy (bottom left panel), and outflow force (bottom right panel) for runs \hrtof\ (pink dashed lines) and \hrtofb\ (purple dot-dashed line) as a function of the primary star's luminosity. The gray solid and dashed lines in the lower right panel indicate the best-fit relationships between force and luminosity for observed massive protostars obtained by \citet{Maud2015b}, and for low-mass protostars by \citet{Bontemps1996a}, respectively.
}
\end{figure*}
As we demonstrated throughout this paper, collimated outflows that originate from accreting protostars dominate the momentum budget for stellar feedback in massive star formation, while radiation pressure is secondary. The momentum from these outflows entrains nearby molecular core material that may eventually be ejected from the core, leading to low SFEs. Studies have observed entrained outflows emanating from MYSOs via methanol masers and molecular line tracers in order to to determine the outflow energetics  (e.g., the entrained outflow mass, momenta, energy, and force) \citep[e.g.,][]{deVilliers2014a, Maud2015b}. These studies have concluded that these quantities scale with the luminosity of the central driving source, indicating that the observed outflows are powered by a common, scalable driving mechanism similar to outflows observed in low-mass star formation \citep{Bontemps1996a, Maud2015b, Bally2016a}. 

The simulations presented in this work allow us to follow the evolution of the the outflow energetics and determine how they depend on source luminosity, and how the relationship between these two quantities in our simulations compares to the observed one. In light of this goal, we show the entrained outflow mass, momenta, energy, and force (i.e., momentum flux) as a function of the massive protostar's luminosity taken directly from runs \hrtof\ (pink dashed lines) and \hrtofb\ (purple dot dashed lines), respectively, in Figure~\ref{fig:OFpropsL}. This figure is constructed to be similar to that presented in \citet{Maud2015b}. We find that the outflow mass, momentum, energy, and force are roughly similar as a function of source luminosity regardless of whether the core is magnetized or not, and that the relationship between outflow force and source luminosity in our simulations agrees well with that determined by \citet{Maud2015b}. 

In the bottom right panel, where we show the outflow force as a function of source luminosity, we overplot the power law fits derived from observations of outflows from low-mass protostars, $\log_{10}F=-5.6+0.9 \times \log_{10}L(L_{\rm \odot})$ \citep[][gray dotted line]{Bontemps1996a}, and massive protostars, $\log_{10}F=-4.8+0.61 \times \log_{10}L(L_{\rm \odot})$ \citep[][gray solid line]{Maud2015b}. We find that the outflow force inferred from our simulations agree well with the fit derived by \citet{Maud2015b}. In their work, \citet{Maud2015b} conclude that the most massive protostars in the clusters are responsible for the energetics of the observed outflows, thereby leading to the shallower slope in their derived fit for the outflow force as compared to the steeper fit determined by \citet{Bontemps1996a} for outflows from low-mass protostars. Our results are consistent with this conclusion.

\section{Conclusions}
\label{sec:conc}
In this work, we performed a series of 3D RHD and RMHD simulations of the gravitational collapse of isolated dense massive prestellar cores to determine how magnetic fields, turbulence, and radiative and collimated outflow feedback affects the formation of massive stellar systems. This is one of the first studies of massive star formation to include all of these effects, in the context of a realistic, turbulent medium without artificially-imposed geometries or symmetries (e.g., stars that are fixed at the origin of a symmetric cloud). By following the material launched by the outflows we have investigated outflow entrainment and the momentum and energy injection by outflows in massive star formation with and without magnetic fields. We reach the following conclusions:

\begin{enumerate}
\item Feedback from outflows dominates over momentum injection by radiation pressure in massive star formation.

\item The accretion and stellar luminosities are comparable throughout the star formation process, therefore accretion luminosity should not be neglected in massive star formation studies.

\item Strong magnetic fields suppress fragmentation and preferentially lead to monolithic collapse to single stars. Weaker magnetic fields and non-ideal effects might weaken this effect, and lead to the formation of a small number of companion stars.

\item The presence of magnetic fields leads to broader and lower density entrained outflows as compared to outflows in non-magnetic cores.

\item Magnetic fields and outflows are required to produce the low SFEs commonly observed in star formation. We find that radiative feedback alone does not lead to SFEs as low as those commonly observed in massive star formation.

\item We find that accretion continues onto the low-mass companion stars even after outflow feedback has significantly reduced accretion onto the massive primary star.

\item  By tracking the launched and entrained outflows we calculated the mass-loading factor and momentum- and energy-scaling factors associated with feedback from protostellar outflows. We find that the mass-loading factor roughly falls between $\sim 2-3$, but that the amounts of momentum and energy carried by the outflow are \textit{smaller} than those injected by the stars, by factors of $\sim 0.25$ and $\sim 0.05$, respectively. Our simulations reproduce the relationship between source luminosity outflow force observed for massive protostars by \citet{Maud2015b}.  

\item Our results for the momentum and energy contained in the entrained outflows suggests that outflows should  be  considered  a  momentum-driven  rather  than an energy-driven feedback mechanism since the reduction in energy is larger than the reduction in momentum.

%\item We use the results of our simulations to propose a sub-grid model to account for protostellar outflows in massive star formation that can be used in future star cluster formation and galaxy formation simulations that are unable to resolve individual star formation.

\end{enumerate}

\software{\textsc{yt}, \citep{Turk2011a}, \orion\ \citep{Li2012a}, \harm\ \citep{Rosen2017a}}

\subsection*{Acknowledgements}
 The authors thank the anonymous referee for their advice and suggestions which improved the manuscript. A.L.R. acknowledges support from NASA through Einstein Postdoctoral Fellowship grant number PF7-180166 awarded by the \textit{Chandra} X-ray Center, which is operated by the Smithsonian Astrophysical Observatory for NASA under contract NAS8-03060. M.R.K. acknowledges support from the Australian Research Council through its \textit{Future Fellowship} and \textit{Discovery Projects} funding schemes (awards FT180100375 and DP190101258), and from the Australian National Computational Infrastructure (NCI, project jh2). A.L.R. would like to thank Stella Offner, Andrew Cunningham, Qizhou Zhang, and Alyssa Goodman for insightful conversations regarding this work. A.L.R. and M.R.K. would also like to thank their ``coworkers,"  Nova Rosen and Jeff Anderson-Krumholz, for ``insightful conversations" and unwavering support at home while this paper was being written during the Coronavirus pandemic of 2020. Their contributions were not sufficient to warrant co-authorship due to excessive napping.\footnote{Because they are cats.}
\bibliographystyle{apj}
%\bibliography{/Users/anna/Dropbox/PostDocLife/refsALR/refs}
%\bibliography{../../../../PostDocLife/refsALR/refs.bib}
\bibliography{refsOF.bib}

\end{document}